\documentclass[aps,preprint,superscriptaddress,showpacs]{revtex4}
\usepackage{epsfig}
\usepackage{natbib}
\usepackage{amsmath}
\usepackage{times}
\usepackage{color}
\usepackage{psfrag}
\usepackage{subfigure}
\begin{document}

\title{
Effective Fokker-Planck Equation for Birhythmic Modified van
der Pol Oscillator}

\author{R. Yamapi}
\email[]{Author to whom correspondence should be addressed. Electronic mail:
ryamapi@yahoo.fr}
\affiliation{Fundamental Physics Laboratory, Department of Physics, Faculty of
 Science,\\
\small  University of Douala, Box 24 157 Douala, Cameroon and \\
\small Salerno unit of CNSIM, Dept. of Physics, Univ. of Salerno,I-84081
Fisciano, Italy}

\author{G. Filatrella}
 \affiliation{Dept. of Sciences for Biological, Geological, and
Environmental Studies\\
\small and Salerno unit of CNSIM, University of Sannio, Via Port'Arsa 11,
I-82100 Benevento, Italy}

 \author{M. A. Aziz-Alaoui }
 \affiliation{ Applied Mathematics Laboratory,
University of Le Havre,\\\small 25 rue ph. Lebon, B.P 540, Le
Havre, Cedex, France}
\author{Hilda A. Cerdeira}
 \affiliation{ Instituto de F\'{i}sica Te\'{o}rica,
\small Universidade Estadual Paulista,
\small Rua Dr. Bento Teobaldo Ferraz, 271,\\
\small Bloco II - Barra Funda, 01140-070 S$\tilde a$o Paulo, Brazil.}

\begin{abstract}

We present an explicit solution based on the phase-amplitude approximation of
the \emph{Fokker-Planck} equation associated with
the \emph{Langevin} equation of the birhythmic modified \emph{van der Pol}
system.
The solution enables us to derive probability distributions analytically
as well as the activation energies  associated to switching between the coexisting different
attractors that characterize the birhythmic system.
Comparing analytical and numerical results we find good agreement when the
frequencies of both attractors are equal, while the predictions of the analytic
estimates deteriorate when the two frequencies depart.
Under the effect of noise the two states that characterize the birhythmic system
can merge, inasmuch as the parameter plane of the birhythmic solutions is found
to shrink when the noise intensity increases. The solution of the
\emph{Fokker-Planck} equation shows that in the birhythmic region,
 the two attractors are characterized by very different probabilities of finding
the system in such a state. The probability becomes comparable only for a narrow
range of the control parameters, thus the two limit cycles have properties in
close analogy with the thermodynamic phases.

\end{abstract}
 \maketitle

{\bf The \emph{van der Pol} oscillator is a model of self-oscillating system
that exhibits periodic oscillations. A modified version -- essentially a higher
order polynomial dissipation -- has been proposed as a model equation for enzyme
dynamics. This model is very interesting as a paradigm for birhythmicity,
it contains multiple stable attractors with different natural frequencies,
therefore it can describe spontaneous switching from
one attractor to another under the influence of noise. The noise induced
transitions between different attractors depend upon the different stability
properties of the attractors, and are usually investigated by means of extensive
\emph{Langevin} simulations. We show that the associated \emph{Fokker-Planck}
equation, in the phase-amplitude approximation, is analytically solvable. The
phase amplitude approximation requires a single frequency, and therefore fails
when the two frequencies of the birhythmic system are significantly different.
However, the approximation is not severe, for it explains the main features of
the system when compared to the numerical simulations of the full model. The
approximated \emph{Fokker-Planck} equation reveals the underlining structure of
an effective potential that separates the different attractors with different frequency, thus explaining the remarkable differences of the stability between the coexisting attractors
that give rise to birhythmicity. Moreover, it reveals that the noise can induce
the stochastic suppression of the bifurcation that leads to birhythmicity.
Finally, the approximated solution shows that the system is located with
overwhelming probability in one attractor, thus being the dominant attractor.
Which attractor is dominant depends upon the external control parameters. This
is in agreement with the general expectation that in bistable systems the
passage from an attractor to the other resembles phase transitions, since only
in a very narrow interval of the external parameters it occurs in both
directions with comparable probabilities.}

\section{Introduction }
\noindent

A stochastic dynamical system is a dynamical system under the effects of noise.
Such effects of fluctuations have been of interest for over a century since the
celebrated work of \emph{Einstein} \cite{1}. Fluctuations are classically
referred to as "noisy" or "stochastic" when their suspected origin implicates
the action of a very large number of variables or degrees of freedom.
For a linear system this leads to the phenomenon of diffusion, while the
coupling of noise to nonlinear deterministic equations can lead to non-trivial
effects \cite{2,3}.
For example, noise can stabilize unstable equilibria and shift bifurcations,
{\it i.e.} the parameter value at which the dynamics changes qualitatively
\cite{4,kar}.
Noise can also lead to transitions between coexisting deterministic stable
states or attractors such as in birhythmic or bistable system
\cite{yamapi-chaos}. Moreover, noise can induce new stable states that have no
deterministic counterpart, for instance noise excites internal modes of
oscillation, and it can even
enhance the response of a nonlinear system to external signals
\cite{5,6}.

In this paper, we investigate analytically the effects of an additive
noise on a special bistable system that displays birhythmicity -- coexisting
attractors that are characterized by different frequencies \cite{h1,h2,h3,h4,h5,h6,h7,h8}. We examine a
birhythmic self-sustained system described by the modified \emph{van-der Pol}
oscillator, subjected to an additive Gaussian white noise \cite{yamapi-chaos}.

Our main aim is to use the phase-amplitude approximation \cite{tutorial}, 
a standard technique for van der Pol \cite{tutorial} and van der Pol - like 
systems \cite{Zakharova}, to derive  an effective \emph{Fokker-Planck} equation 
\cite{risken} that can be analytically managed. This allows us to 
analytically derive the activation energies associated to the switching
between different attractors \cite{yamapi-chaos,chamgoue-epjb}. The analytical solution of
the approximated model is not limited to vanishingly small noise intensity as it
was done for the numerical estimate of the escape time \cite{yamapi-chaos} to
derive the pseudopotential \cite{Graham85}. Another purpose of the present paper is to verify, with numerical simulations, that in spite of the approximations the analytical probability
distribution is reliable.

The paper is organized as follows: Section II presents the modified
\emph{van-der Pol} system with an additive Gaussian white noise. Section III
deals with the derivation and analysis of an effective \emph{Fokker-Planck}
equation for the birhythmic modified \emph{van der Pol} oscillator. The
probability distribution given by the approximated \emph{Fokker-Planck} equation
is analyzed and the activation energies are derived. In Section IV, we integrate
numerically the
stochastic second order differential equation and discuss the results. Section V
concludes.

\section{The birhythmic properties of the noisy model}
\subsection{The modified  van der Pol oscillator with an additive noise}
\noindent

The model considered is a  \emph{van der Pol} oscillator with a nonlinear
dissipation of higher polynomial order described by the equation (overdots as
usual stand for the derivative with respect to time)
\begin{eqnarray}
\label{eq1} \ddot x-\mu (1- x^2+\alpha x^4-\beta x^6)\dot x+x=0.
\end{eqnarray}
This model was proposed by Kaiser \cite{kaiserv} as more appropriate than the
\emph{van der Pol} oscillator to describe certain specific processes in
biophysical systems. In fact the modified \emph{van der Pol}-like oscillator
described by Eq.~(\ref{eq1}) is used to model coherent oscillations in
biological systems, such as an enzymatic substrate reaction with ferroelectric
behavior in brain waves models (see Refs.
\cite{ecyw,kaiser1,frohlich,enjieu-yamapi-chabi} for more details). From the
standpoint of nonlinear dynamics, it represents a model which exhibits an
extremely rich bifurcation behavior.
The quantities $\alpha$ and $\beta$ are positive parameters which measure the
degree of tendency of the system to a ferroelectric instability compared to its
electric resistance, while $\mu$ is the parameter that tunes nonlinearity
\cite{ecyw}.
The model Eq.(\ref{eq1}) is a nonlinear self-sustained oscillator which
possesses more than one stable limit-cycle solution \cite{kaiser2}. Such systems
are of interest especially in biology, for example to describe the coexistence
of two stable oscillatory states, as in
enzyme reactions \cite{li}. Another example is the explanation of
the existence of multiple frequency and intensity windows in the
reaction of biological systems when they are irradiated with very
weak electromagnetic fields
\cite{kaiser1,kaiser2,kaiser3,kaiser4,kaiser5,kaiser7}.
Moreover, the model under consideration offers general aspects
concerning the behavior of nonlinear dynamical systems. \emph{Kaiser} and
\emph{Eichwald} \cite{kaiser7} have analyzed the super-harmonic
resonance structure, while \emph{Eichwald} and \emph{Kaiser} \cite{kaiserv} have
found symmetry-breaking crisis and intermittency.

In Ref. \cite{ecyw} an analytical approximation has been derived for the
coexisting oscillations of the two attractors with different natural frequencies
for the deterministic part of the model equation. A numerical investigation of
the escape times (and hence of the activation energies) has suggested that the
stability properties of the attractors can be very different
\cite{yamapi-chaos}. It has further been shown that time delayed feedback leads
to stabilization \cite{Ghosh11}, also in the presence of external noise
\cite{chamgoue-epjb}.

Noise can enter the system for instance, through the electric field applied to
the excited enzymes which depends on the external chemical influences or through
the flow of enzyme molecules. One can therefore assume that the environmental
influence contains a random perturbation and to postulate that the activated
enzymes are subject to a random excitation governed by the \emph{Langevin}
version of Eq.~(\ref{eq1}), namely:
\begin{eqnarray}
\label{eq2}
\ddot x-\mu (1 - x^2 +\alpha x^4 -\beta x^6)\dot x + x = \Gamma(t).
\end{eqnarray}
\noindent $\Gamma(t)$ can be assumed to be an   additive Gaussian white noise
with arbitrary amplitude $D$ \cite{tutorial} and it has the properties:
\begin{eqnarray}
\label{eq3}
&&<\Gamma(t)>=0\nonumber \\
&&<\Gamma (t),\Gamma (t')>=2D\delta (t-t')
\end{eqnarray}

\noindent which completely determine its statistical features. The noise term is here treated as external \cite{Hanggi82}, i.e. due to a disturbance from the environment and not subject to the fluctuation dissipation theorem.

\subsection{Birhythmic properties} \noindent

Without noise ($\Gamma=0$), Eq.(\ref{eq2}) reduces to the modified version of
the \emph{van der Pol} oscillator (\ref{eq1}) which has steady-state solutions
that depend on the parameters $\alpha, \beta$ and $\mu$ and correspond to
attractors in state space. The dynamical attractors of the free-noise modified
\emph{van der Pol} Eq.(\ref{eq1}) have been  determined  analytically, the
expressions of the amplitudes $A_i$ and frequency $\Omega_i$ (i=1,2,3) of the
limit-cycle solutions have been established in
Ref.\cite{chamgoue-epjb,ecyw,yamapi-chaos}, in which the periodic solutions of
the modified  \emph{van der Pol} oscillator (\ref{eq1}) are approximated by
\begin{eqnarray}
\label{eq6}
x(t)=A\cos \Omega t .
\end{eqnarray}
The amplitude $A$ is independent of the coefficient $\mu$ up to corrections of
the order $\mu^2$ and implicitly given by the relation:
\begin{eqnarray}
\label{eq7}
\frac{5\beta}{64}A^6-\frac{\alpha}{8}A^4+\frac{1}{4}A^2-1=0.
\end{eqnarray}

\noindent The coefficient $\mu$ enters in the expression for the frequency
$\Omega$ as a second order correction:
\begin{eqnarray}
\label{eq8} \Omega=1+\mu^2\omega_2+o(\mu^3)
\end{eqnarray}

\noindent thus the deviations of the frequency from the linear harmonic solution
are characterized by an amplitude dependent frequency
\cite{enjieu-yamapi-chabi}:
\begin{equation}
\omega_2=\frac{93\beta^2}{65536}A^{12}-\frac{69\alpha\beta}{16384}A^{10}+(\frac{
67\beta}{8192}+
\frac{3\alpha^2}{1024})A^8-(\frac{73\beta}{2048}+\frac{\alpha}{96})A^6+(\frac{1}
{128}+
\frac{\alpha}{24})A^4-\frac{3}{64}A^2
\label{eq:amplitude}
\end{equation}

\noindent  Depending on the values of the parameters $\alpha$ and $\beta$, the
modified \emph{van der Pol} oscillator possesses one or three limit cycles. In
fact, Eq.(\ref{eq7}) can give rise to one or three positive real roots that
correspond to one stable limit cycle or three limit cycle solutions (of which
two are stable and one is unstable), respectively.
The dynamical attractors and birhythmicity ({\it i.e.} the coexistence between
two stable regimes of limit cycle oscillations) are numerically found solving
the amplitude Eq.(\ref{eq:amplitude}) \cite{yamapi-chaos}. The three roots
$A_1$, $A_2$, and $A_3$ denote the inner stable orbit, the unstable orbit, and
the outer stable orbit, respectively.
When three limit cycles are obtained, Eq.(\ref{eq8}) supplies the frequencies
$\Omega_{1,2,3}$ in correspondence of the roots $A_{1,2,3}$.
Being one of the attractors unstable, the system only displays two frequencies
$\Omega_{1,3}$ (and hence birhythmicity) at two different amplitudes $A_{1,3}$,
while the unstable limit cycle of amplitude $A_2$ represents the separatrix
between the basins of attraction of the stable limit cycles.
We show in Fig.1 the region of existence of birhythmicity
in the two parameters phase space ($\alpha$-$\beta$)
\cite{ecyw,enjieu-yamapi-chabi} (the two coexisting stable limit
cycle attractors can be found in \cite{yamapi-chaos}). The
question we want to address is the influence of noise on the above
properties investigating the response of an additive Gaussian white noise in the
phase-amplitude limit. In Ref.\cite{yamapi-chaos} the system has been
numerically tackled in the regime of vanishingly small noise. In this limit the
escape rate gives an effective potential that acts as an activation barrier. We
employ the phase-amplitude approximation that should be both faster (being
analytical) and more accurate at finite values of the noise, as will be
discussed in the next Sect. $III$.

\section{Analytical estimates}
\noindent

The analytic results on the deterministic system are based on the approximated
cycle given by Eq.(\ref{eq6}). Quite naturally, one can treat the noise in the
system starting from such approximation. To this extent, we rewrite the
\emph{Langevin} Eq.(\ref{eq2}) in a system of two coupled first order
differential equations:
\begin{eqnarray}
\label{eq9}
\dot x&=&u, \nonumber \\
\dot u&=&\mu(1-x^2+\alpha x^4-\beta x^6)u-x +\Gamma.
\end{eqnarray}

\noindent We seek for solution in the context of the phase-amplitude
approximation, {\it i.e.}
 letting the amplitude and the phase of Eq.(\ref{eq6}) to be time dependent
\cite{tutorial}:
\begin{eqnarray}
\label{eq10}
x&=& A(t)\cos(\Omega t+\phi(t)) \nonumber \\
u&=&-A(t)\omega_0\sin(\Omega t+\phi(t)).
\end{eqnarray}

\noindent Inserting Eq.(\ref{eq10}) into Eq.(\ref{eq9}), one retrieves a system
of two \emph{Langevin} equations for the amplitude $A(t)$ and phase $\phi(t)$
variables, that is, of course, as difficult to manage as the original model
(\ref{eq9}). We will follow the standard analysis of nonlinear oscillators
\cite{GF,Zakharova} that consists in assuming that in a period $2\pi/\Omega$ the
variables $A(t)$ and $\phi (t)$ do not change significantly, so one can average
the effect of the random perturbation \cite{chamgoue-epjb}. Although in
principle this method also relies on the smallness of the noise, since the
averaging requires that the approximate solution (\ref{eq10}) is not
significantly altered in a cycle $2\pi/\Omega$, the procedure has proven very
robust in a similar \emph{van der Pol - Duffing} oscillator \cite{Zakharova}.
It is important to note that for $\mu=0$ the system reduces to the harmonic oscillator, as described by the solution Eq.(\ref{eq10}) with constant amplitude and phase. Since we are interested in the influence of noise $D$ and nonlinear dissipation ($\alpha$ and $\beta$) in birhythmic systems, we keep the parameter $\mu$ small ($\mu =0.1$). If the present model is employed to model the population of enzyme molecules, the parameter represents the difference between the thermal activated polarization and the external field induced polarization \cite{kaiser2}.
However, we note that for a birhythmic system a further difficulty occurs: the
system has two different frequencies $\Omega_1\neq \Omega_3$, while the
approximation (\ref{eq10}) is monorhythmical. Assuming that the two frequencies
are not too different, we insert Eq.(\ref{eq10}) into Eq.(\ref{eq9}) and
average, to retrieve the effective (and simpler) \emph{Langevin} equation for
the amplitude $A$ and phase $\phi$ variables:
\begin{eqnarray}
\label{eq11}
\dot \phi&=&-\frac{\Omega^2-1}{2\Omega}-
\sqrt{\frac{D}{2A\Omega^2}}<\Gamma(t)>,
\nonumber \\
\dot A&=&\frac{\mu A}{2}[(1-\frac{1}{4}A^2+\frac{1}{8}\alpha
A^4-\frac{5}{64}\beta A^6)]-\sqrt{\frac{D}{2\omega_0^2}}<\Gamma(t)>.
\end{eqnarray}

\noindent We thus study the system in the slow averaged variables; for the slow variables the average noise can still be considered white and uncorrelated \cite{tutorial}, and the \emph{Fokker-Planck} equation associated to the \emph{Langevin}
model (\ref{eq11}) reads:
\begin{eqnarray}
\label{eq12} \frac{\partial P}{\partial t}=-\frac{\partial
S^{\phi}}{\partial \phi} - \frac{\partial S^{A}}{\partial A},
\end{eqnarray}
where $S=S^{\phi}+S^{A}$ is the probability current defined by:
\begin{eqnarray}
\label{eq13}
S^{\phi}&=& K_1^{\phi}P-\frac{\partial }{\partial \phi}(K_2^{\phi,\phi}P),
\nonumber\\
S^{A}   &=& K_1^{A}P   -\frac{\partial }{\partial A}(K_2^{A,A}P).
\end{eqnarray}

\noindent The drift coefficients $K_1^{\phi}$ and $K_1^{A}$ associated to
Eq.(\ref{eq11}) read:
\begin{eqnarray}
\label{eq14}
K_1^{\phi} &=& -\frac{\Omega^2 }{2\Omega} \nonumber \\
K_1^{A} &=& \frac{\mu A}{2} [1-\frac{1}{4}A^2+\frac{1}{8}\alpha
A^4-\frac{5}{64}\beta A^6]+\frac{D}{2\Omega^2A}.
\end{eqnarray}

\noindent The off diagonal diffusion coefficients $K_2^{\phi,A}$ and
$K_2^{A,\phi}$ vanish, while the diagonal coefficient read:
\begin{eqnarray}
\label{eq15}
K_2^{\phi,\phi} &=& \frac{D}{(\Omega A)^2}, \nonumber \\
K_2^{A,A}       &=& \frac{D}{2\Omega^2}.
\end{eqnarray}

\noindent We seek for stationary solutions, $\partial P/\partial t =0$ of
Eq.(\ref{eq12}). We note that in the averaged equation (\ref{eq10}) for the phase $A$ the phase $\phi$ does not appear, and therefore the integration over all phases gives rises to a normalization constant. We
therefore only seek solutions for the probability distribution associated to the
constant probability current, $S^{A} =
const$. Moreover, since the probability distribution must vanish for
$A=\infty$, we can set the constant to $0$. Finally, the equation for the radial
part of the probability distribution $P$ reads:
\begin{eqnarray}
\label{eq16} S^{A} = 0 \Rightarrow K_1^{A}P = \frac{d }{d A}(K_2^{A,A}P),
\end{eqnarray}

\noindent or, explicitly:
\begin{eqnarray}
\label{eq17}
P(A) = c A \exp \{ \frac{\mu \Omega^2}{2D}A^2
[1-\frac{1}{8}A^2+\frac{1}{24}\alpha
A^4-\frac{5}{256}\beta A^6 ] \},
\end{eqnarray}

\noindent where $c$ is a constant of normalization. This solution contains as
 particular cases the harmonic oscillator ($\mu<0$, $\alpha=\beta=0$, and
discarding the $A^2/8$ term) and the standard \emph{van der Pol} oscillator
($\mu>0$, $\alpha=\beta=0$).

The probability distribution is in general very asymmetric, for most of the
parameters $\alpha$ or $\beta$ one  can localize the probability function around
a single orbit. Before proceeding further in our analysis, it should be noted
that the peaks of the probability distribution can be located using the
following equation:
\begin{equation}
\label{eq18}
\frac{d log(P)}{dA}=0\Longrightarrow
\left [
\frac{5}{64}\beta A^6-\frac{1}{8}\alpha A^4+\frac{1}{4}A^2-1
\right ]
A^2-\frac{D}{\mu \Omega^2}=0.
\end{equation}

\noindent For $D = 0$, the amplitude (\ref{eq18}) coincides with the
deterministic amplitude equation (5)\cite{chamgoue-epjb}. In Fig.2  we report
the influence of the noise
intensity $D$ on the region of multi-limit cycle orbits of Fig. 1. In the
parametric $(\alpha,\beta)$-plane of Fig. 2 it is evident the effect of the
noise intensity $D$ on the transition boundary between the appearance of single
and multi-limit cycles orbits: the bifurcation that leads to birhythmicity is
postponed under the influence of noise \cite{Lefever}. As a consequence, the
region of existence of three limit cycles, a condition for birhythmicity,
decreases with the increase of the noise intensity and disappears altogether for
high noise intensity.

An important feature of birhythmicity in the present model is highlighted in
Fig. 3. We first define ${\cal P}_{1,3}$ of the probability to find
the system in the basin of attraction of each stable orbit $1$ and $3$:
\begin{eqnarray}
{\cal P}_{1} &=& \int_0^{A_2}P(A)dA, \nonumber \\
{\cal P}_{3} &=& \int_{A_2}^{\infty}P(A)dA.
\label{eq:Tdefinition}
\end{eqnarray}

\noindent These quantities measure the relative stabilities pertaining to the
attractors $1$ and $3$ and are related to the resident time by the relation ${\cal P}_{1,3}=T_{1.3}/(T_1+T_3)$, so that ${\cal P}_1/{\cal P}_3=T_1/T_3$. In Fig.\ref{fig2a} we show in the parameter plane $\alpha-\beta$ the locus where the system stays with equal probability on both
attractors (solid line) $T_1=T_3$.
We also show two further curves: the limit where the first attractor is much
more stable than the other ($T_1 \ge 10T_3$, circles) and the passage to the
reverse situation ($T_3 \ge 10 T_1$, crosses). From the figure it is evident
that the outer attractor is dominantly visited in most of the parameter plane.
Moreover, the transition from the two opposite cases (\emph{i.e.}, a change of
two order of magnitudes of the relative resident times) occurs with a very
narrow change of the control parameters $\alpha$ and $\beta$. The drastic change
is further investigated in Fig. 4, where we show a blow-up of the crossover
region around $T_1=T_3$ for different values of the parameter $\beta$. The
$\alpha$ value is increased up to the maximum value when birhythmicity
disappears. The general behavior observed for all $\beta$ values, closely
reminds phase transitions: the probability to find the system in one condition
(around the attractor $A_1$) or the other (around the attractor $A_3$)
drastically changes in a very small interval of the $\alpha$ parameter. The same
behavior, this time with a constant value of $\alpha$ and varying $\beta$ is
shown in Fig. 5. The effect of the noise intensity is much less pronounced, see
Fig. 6. It is apparent that the effective temperature is capable to cause a
crossover between the residence times only in a very narrow region of the phase
space, inasmuch as the noise causes a contraction of the region of existence of
birhythmicity. However it is evident that the transition is much slower, and a
crossover only occurs in the limited parameter space around $\alpha=0.05$,
$\beta=0.0005$.

The stability properties of the two attractors have also been investigated in
the limit of small noise values \cite{yamapi-chaos}, where it has been found the
same asymmetrical behavior of the probability distribution, with a
sudden change for small variations of $\alpha$ and $\beta$. In fact one can
notice that the effective \emph{Langevin} equation
(\ref{eq11}) amounts to the Brownian motion of a particle
in a double well, whose potential reads \cite{chamgoue-epjb}:
\begin{eqnarray}
\label{eq19}
\dot A&=&-\frac{\partial F_A(A)}{\partial A} -\sqrt{\frac{D}{2w_0^2}}<\Gamma(t)> ,
\nonumber \\
F_A(A)&=& -\frac{\mu A^2}{4}[(1-\frac{1}{8}A^2+\frac{1}{24}\alpha
A^4-\frac{5}{256}\beta A^6)].
\end{eqnarray}
It is therefore evident that the transition from the inner orbit
$A_1$ to the outer orbit $A_3$ through the unstable orbit $A_2$,
as well as the inverse process, can be interpreted as the diffusion over an
effective potential barrier, and therefore the escape times are given by
the \emph{Kramer's} inverse rate \cite{Hanggi82}, for instance used in Josephson physics to detect the classical quantum/classical crossover \cite{longobardi11} or for signal detection \cite{addesso12}.
\begin{eqnarray}
\label{eq20}
\tau_{1\rightarrow 3} &\propto& \exp\left[\frac{4\Omega^2}{D}\left(
F_A(A_2)-F_A(A_1) \right) \right]=
\exp \left[\frac{\Delta U_1}{D} \right] \nonumber \\
\tau_{3\rightarrow 1} &\propto& \exp
\left[\frac{4\Omega^2}{D}\left( F_A(A_2)-F_A(A_3) \right)
\right]=  \exp \left[\frac{\Delta U_3}{D} \right].
\end{eqnarray}
Thus the average time to pass from one attractor to the other is analogous to the passage over a barrier.
The pseudopotential barrier numerically derived in Ref.\cite{yamapi-chaos} is
therefore, in the phase-amplitude approximation, an effective
potential  for the amplitude variables \cite{chamgoue-epjb}. Since the effective
potential is analytical, we can confirm several features of the
pseudo-potential, for instance that  the potential barriers are
proportional to the nonlinear parameter $\mu$ \cite{yamapi-chaos}. It is also
interesting to  investigate the behavior of the potential barriers of
Eq.(\ref{eq20}) as a function of the parameters $\alpha$ an $\beta$, the
analogous of the analysis of Eq.(\ref{eq17}) in Figs. 3,4,5,6. Inspection of the
effective potential (\ref{eq20}), confirms that it is very asymmetrical, since
one energy barrier is generally much higher than the other. Combining this
observation with the exponential behavior of the escape rates (\ref{eq20}) one
deduces that the system does not equally stays on both attractors, but rather it
clearly exhibits a preference for one attractor with respect to the other (the
relative occupancies  read $T_1/T_3 \simeq \exp[(\Delta U_3-\Delta U_1)/D]$
\cite{Hanggi82}). One concludes that the birhythmic system behaves as a bistable
tunnel diode \cite{Hanggi82}: keeping fixed a control parameter (say $\beta$)
and changing the other ($\alpha$ in this case) the weight of the probability
distribution is concentrated in the proximity of one or the other of the two
stable deterministic solutions of Eq.(\ref{eq1}), thus obtaining again a first
order phase transition. This result supports the notion that the analogy with
phase transitions is generic for bistable oscillators \cite{Stambaugh06}.

\section{Numerical simulations and Results}

\noindent  To check the validity of the approximations behind the analytic
treatment that has led to the solution (\ref{eq17}),
 we have performed numerical simulations of the Langevin dynamics (\ref{eq2}).
There are several methods and algorithms for solving second-order
stochastic differential equations \cite{Burrage07} as the implicit midpoint rule
with {\it Heun} and {\it Leapfrog} methods  or faster numerical algorithms such
as the stochastic version of the
{\it Runge-Kutta} methods and a quasisymplectic algorithm \cite{Mannella04}. To
prove that the simple procedure given by the {\it Euler} algorithm is reliable,
we have employed it in a few selected points with two different methods. The
starting point is the {\it Box-Mueller} algorithm \cite{knuth} to generate a
Gaussian white noise distributed random variable $\Gamma_{\Delta t}$ from two
random numbers $a$ and $b$ which are uniformly distributed on the unit interval
$[0,1]$. The random number approximates the effect of the noise of intensity $D$
over the interval $\Delta t$ in the {\it Euler} algorithm for the integration of
Eq.(\ref{eq9}). We have then halved the step
size $\Delta t$ until the results became independent of the step size; the step
size used for all numerical integration is $\Delta t=0.001$. To verify the
numerical results obtained with the Euler method, we have used a
quasi-symplectic algorithm of \emph{Mannella} \cite{Mannella04} to numerically
compute the probability distribution. The logic behind the choice to compare the
\emph{Euler} algorithm with a quasi-symplectic algorithm is that the nonlinear
dissipation of the model (\ref{eq2}) oscillates and vanishes twice in each
cycle. We have therefore checked the results with an algorithm that has proved
to perform independently of the dissipation value \cite{Burrage07}.

In Fig. 7 we plot the behavior of the probability
distribution $P$  as a function of the amplitude $A$ for several values of
 the noise intensity $D$, when the frequencies of both attractors are similar,
\emph{i.e.} $\Omega_1\simeq
\Omega_3\simeq 1$.
It clearly shows that the system is more likely found
at two distinct distances from the origin, the essential feature
of birhythmicity. In general, for the set of
parameters $\alpha=0.083,\beta=0.0014$, the probability distribution $P$ is
asymmetric. As observed in Sec. $III$ the probability distribution changes with
a small variation of the parameters $\alpha$ and $\beta$
\cite{yamapi-chaos,chamgoue-epjb}.
It is important to note that the agreement between numerical and analytical
results is fairly good for low $A$ values around the inner orbit, when the
frequency of one attractor is very similar to $1$, while for larger amplitude
$A$ the agreement becomes progressively poorer. However, it seems that the
phase-amplitude approximation is capable to capture the main feature of the
phenomenon: an increase or a decrease of the amplitude when the fluctuation
parameter $D$ is varied. At high fluctuations ($D>1$) the system becomes
monorhythmical, see also Fig. 2, thus confirming the noise induced transition
from bimodal to unimodal, sometimes referred to as phenomenological bifurcations
\cite{Zakharova}.

As mentioned in Sec. $III$, the phase-amplitude approximation is not appropriate
when the two frequencies of the attractors are different, \emph{i.e.}
$\Omega_1\neq
\Omega_3$. In fact numerical simulations in this case show a poor agreement, see
Fig. 8 where $\Omega_1\simeq 1$ and $\Omega_3 \simeq 0.8$. This shows the
limitations of this analysis
of phase-amplitude approximation.

Let us return to Eq.(\ref{eq18}) that shows how the orbits radii depend on the
noise intensity $D$. The analytical and numerical behaviors of the limit cycle
attractors are reported in Figs. 9 and 10 that show amplitudes $A_{1,3}$ and
the  associated bandwidths $\Delta A_{1,3}$ (the width when the height of the
probability peaks is reduced of a factor $2$) as a function of the noise
intensity $D$ for two sets of parameters $\alpha$ and $\beta$. We find that the
amplitudes $A_1$ and $A_3$ change very slightly when the noise intensity
increases. Also the bandwidth slightly increases with the noise intensity $D$.
Through Eqs. (\ref{eq17},\ref{eq18}) one can derive the behavior of the
effective potential barriers \cite{yamapi-chaos,chamgoue-epjb}. We have
numerically compared the analytic predictions with simulations in Fig. 11, where
we plot $\Delta U_1$ and $\Delta U_3$
as a function of the parameter $\alpha$ with $\beta=0.002$. It should be noted
that according to the numerical results of Ref.\cite{yamapi-chaos}, varying
$\alpha$  from $\alpha=0.095$ to $\alpha=0.135$ the system passes from the
region where $\Omega_1 \simeq \Omega_3$ to the region with
$\Omega_1\neq\Omega_3$.
It is clear that in general the two energy barriers are very different. For low
$\alpha$ values $\Delta U_3$ is well approximated by the analytic approach,
while for larger $\alpha$ the agreement becomes progressively poorer.
Nevertheless it seems that the phase-amplitude approximation is capable to
capture the main feature of the phenomenon: an increase or a decrease of the
activation energies when the dissipation parameters are varied.
An analogous behavior is observed in Fig. 12, where we
plot the behaviors of $\Delta U_1$ and $\Delta U_3$ as a function of
the parameter $\beta$.

\section{Conclusions}

We have approached a theoretical description of the temporal evolution of
the  modified  \emph{van der Pol} oscillator with an additive Gaussian white
noise  in the region where birhythmicity (in the absence of noise) occurs. To get an analytical insight on this system we have used an explicit solution based on the phase-amplitude approximation of the \emph{Fokker-Planck} equation to analytically derive the probability distributions.
The activation energies associated to the switches between different attractors
have been derived analytically and numerically.
We have found that the agreement is fairly good.
The characteristics of the birhythmic properties in a modified \emph{van der
Pol}
 oscillator are strongly influenced by both the nonlinear coefficients $\alpha$,
$\beta$ and the noise intensity $D$.
The boundary of the existence of multi-limit-cycle solutions,
in the parametric $(\alpha,\beta)$-plane, decreases with the increase of the
noise intensity D.
 Finally, the analytic estimate of the stability of the two attractors varies
with the control parameters (the dissipation $\alpha$ and $\beta$) in a way that
resembles phase transitions: for most parameters value the system is located
around only one attractor, the other being visited with a vanishingly small
probability. Only at special values of the control parameters the residence
times are comparable, in agreement with experimental observations of
birhythmicity in Biological systems: the passage from an attractor to another
only occurs by varying the external parameters and not under the influence of
noise \cite{Hounsgaard98,Geva06}.
\vspace{-0.5cm}
\section*{Acknowledgments}
\noindent

\footnotesize{\emph{R.Yamapi} undertook this work with the support of the
ICTP (International Centre for Theoretical Physics) in the
framework of Training and Research in Italian Laboratories (TRIL) for AFRICA
programme, Trieste, Italy and the CNPq-ProAfrica project nr. 490265/2010-3 (Brazil). He also acknowledges the hospitality of the
Dipartimento di Fisica "E.R. Caianiello" of the Universit\`a di Salerno,
Fisciano, Italy and the Institute of Theoretical Physics, UNESP, S$\tilde a$o Paulo, Brazil.
}

\newpage

\begin{figure}[htb]
\centering
\begin{minipage}{12cm}
\begin{center}
\begin{picture}(150,70)
\put(0,0.0){\psfig{file=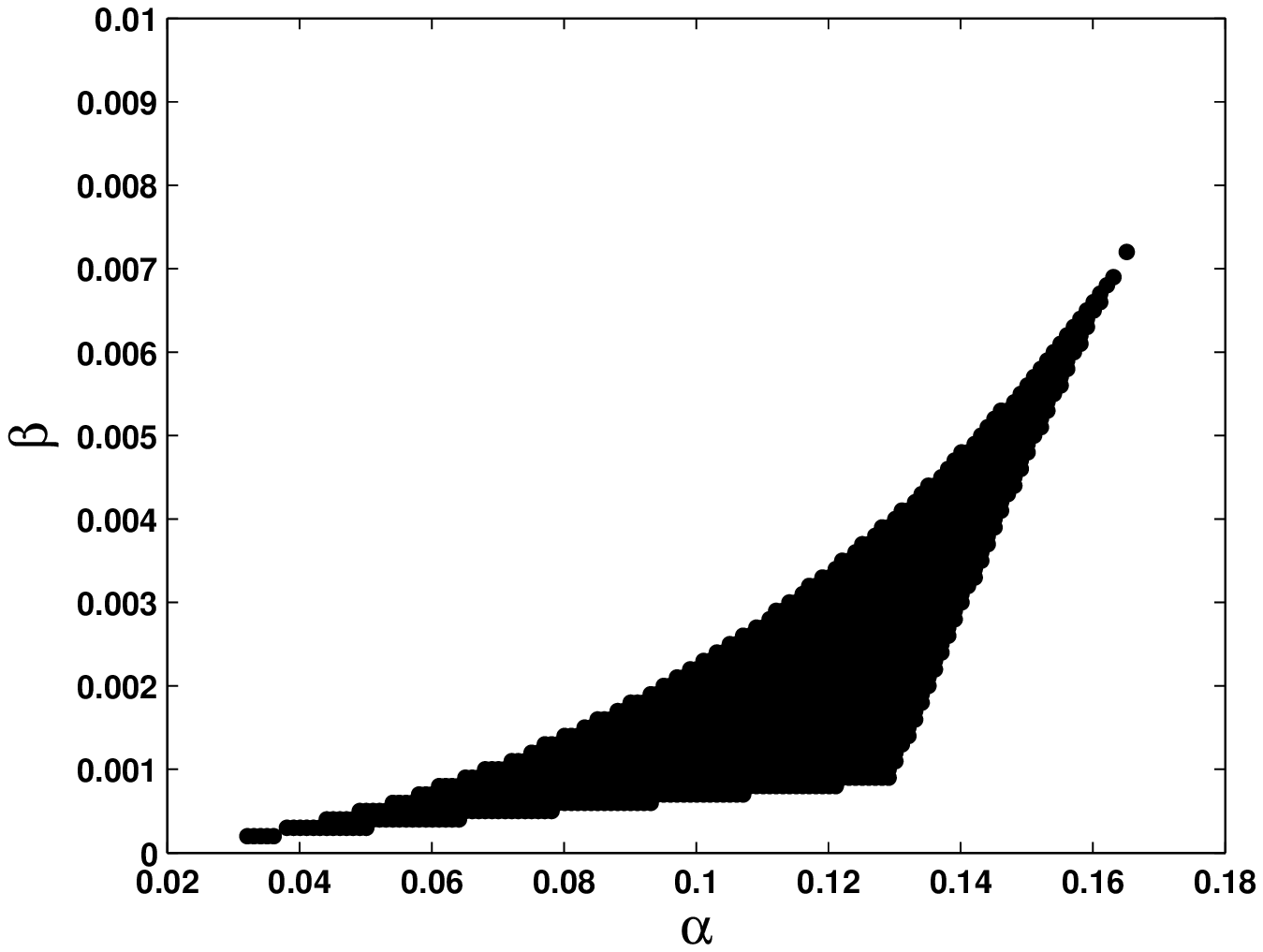,width=12cm,height=7cm,angle=0.0}}
\end{picture}
\caption[] {\footnotesize \it Parameters domain for the existence
of a single limit cycle (\emph{white area}) and three limit cycles (\emph{black
area}) solutions of Eq. (\ref{eq1}) for $\mu=0.1$.}
\label{fig1}
\end{center}
\end{minipage}
\end{figure}

\newpage

\begin{figure}[htb]
\centering
\begin{minipage}{12cm}
\begin{center}
\begin{picture}(150,70)
\put(0,0.0){\psfig{file=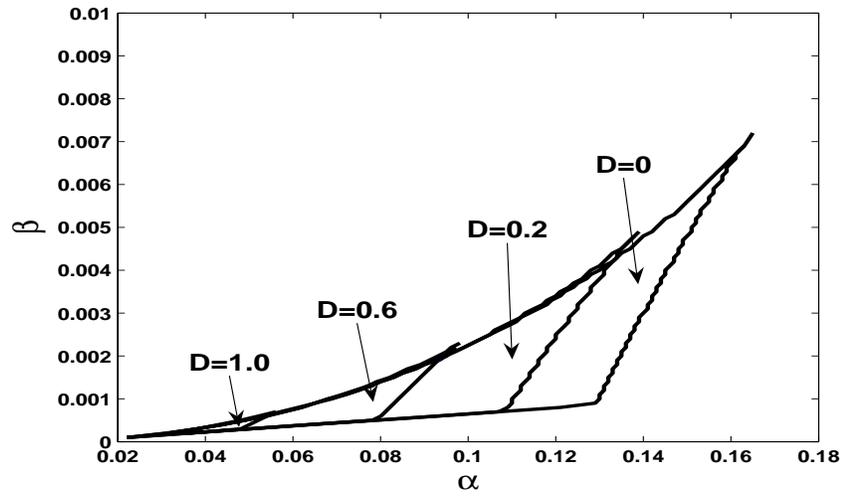,width=12cm,height=7cm,angle=0.0}}
\end{picture}
\caption[] {\footnotesize \it Effect of the noise intensity $D$ on
the boundary between the region of one and three limit-cycle
solutions in the parametric $(\alpha,\beta)$-plane of the {\emph Fokker-Planck}
Eq.(\ref{eq12})) for $\mu = 0.1$ as in Fig.\ref{fig1}. }
\label{fig2}
\end{center}
\end{minipage}
\end{figure}

\begin{figure}[htb]
\centering
\begin{minipage}{12cm}
\begin{center}
\begin{picture}(150,70)
\put(0,0.0){\psfig{file=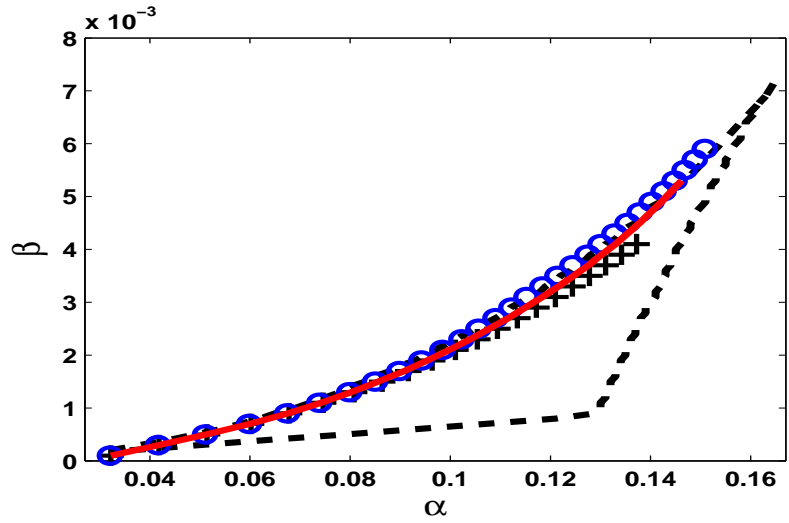,width=12cm,height=7cm,angle=0.0}}
\end{picture}
\caption[] {\footnotesize \it Behavior of the residence times in the parameter
space. The solid line denotes the locus $T_1=T_3$, while circles and crosses
denote the situation where $T_1=10T_3$ and $T_1=T_3/10$, respectively. The
dashed line is the border of existence of birhythmicity. The noise level is
$D=0.1$ }
\label{fig2a}
\end{center}
\end{minipage}
\end{figure}

\begin{figure}[htb]
\centering
\begin{minipage}[10,10]{15cm}
\begin{center}
\begin{picture}(200,160)
\put(20,110){\psfig{file=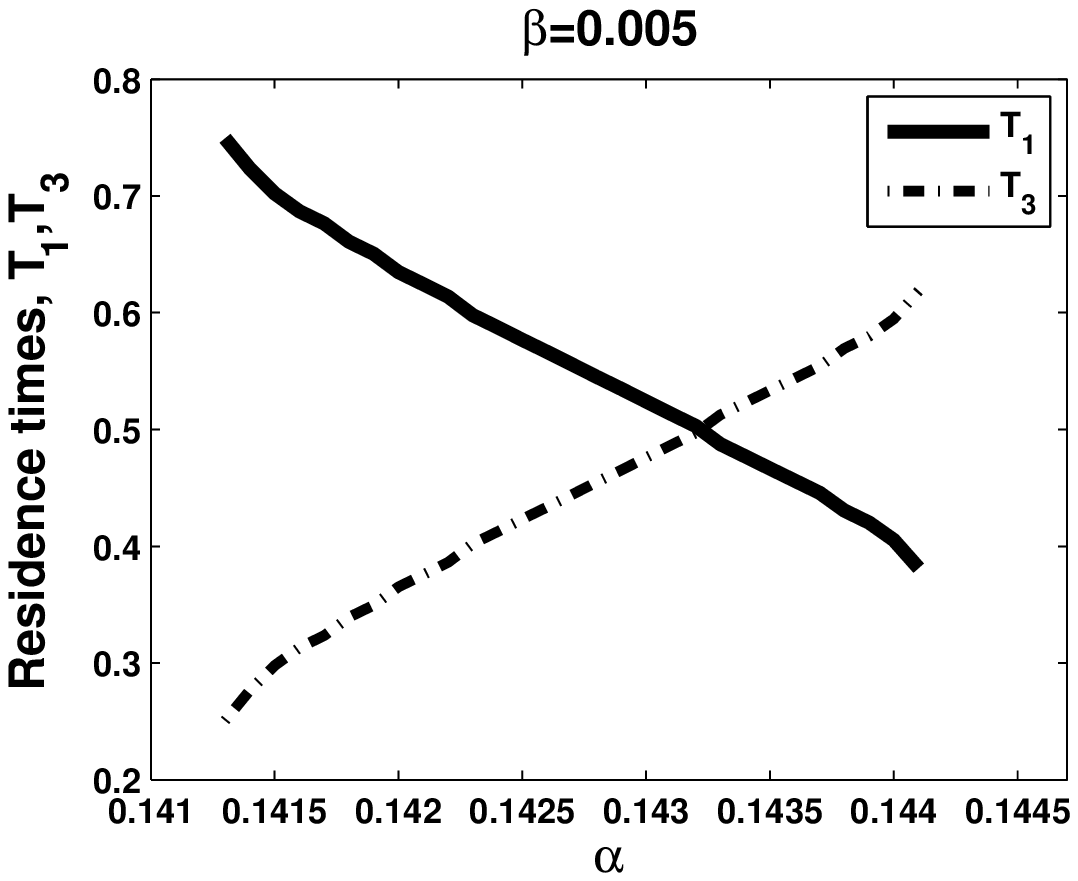,width=10cm,height=5cm,angle=0.0}}
\put(20,55){\psfig{file=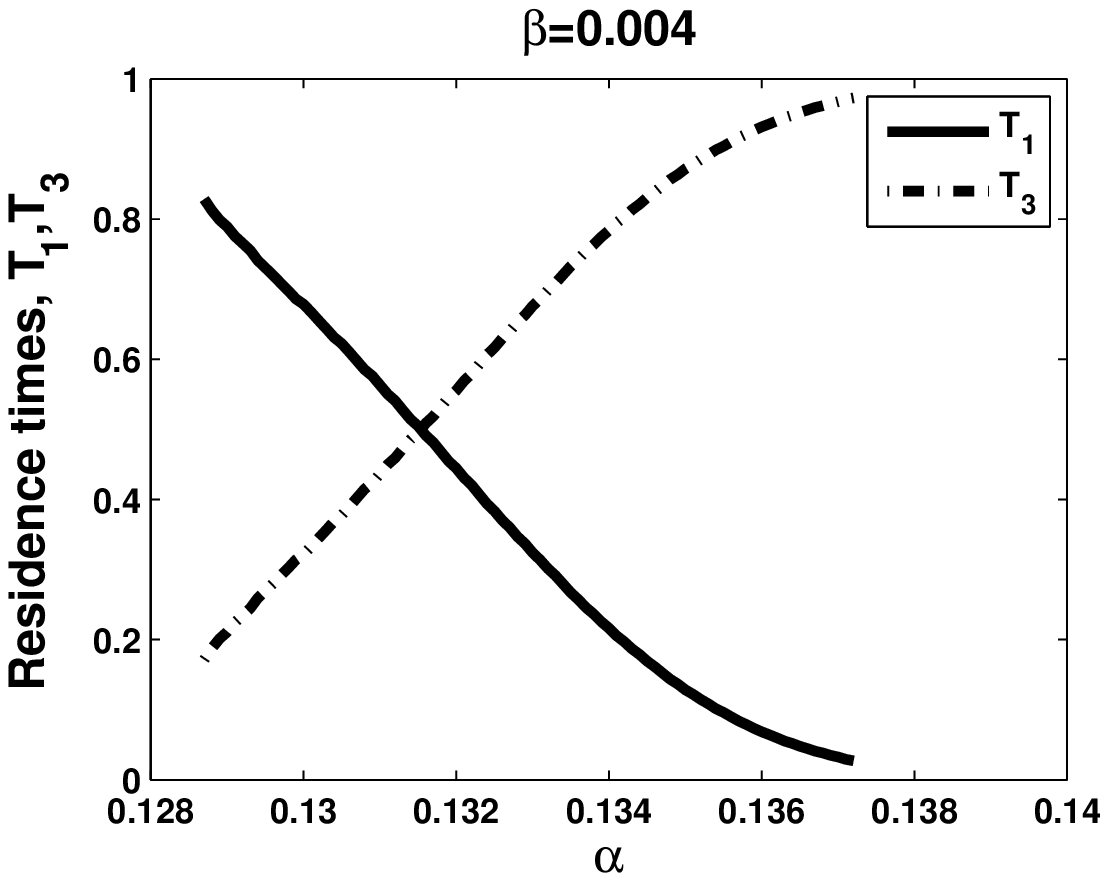,width=10cm,height=5cm,angle=0.0}}
\put(20,0){\psfig{file=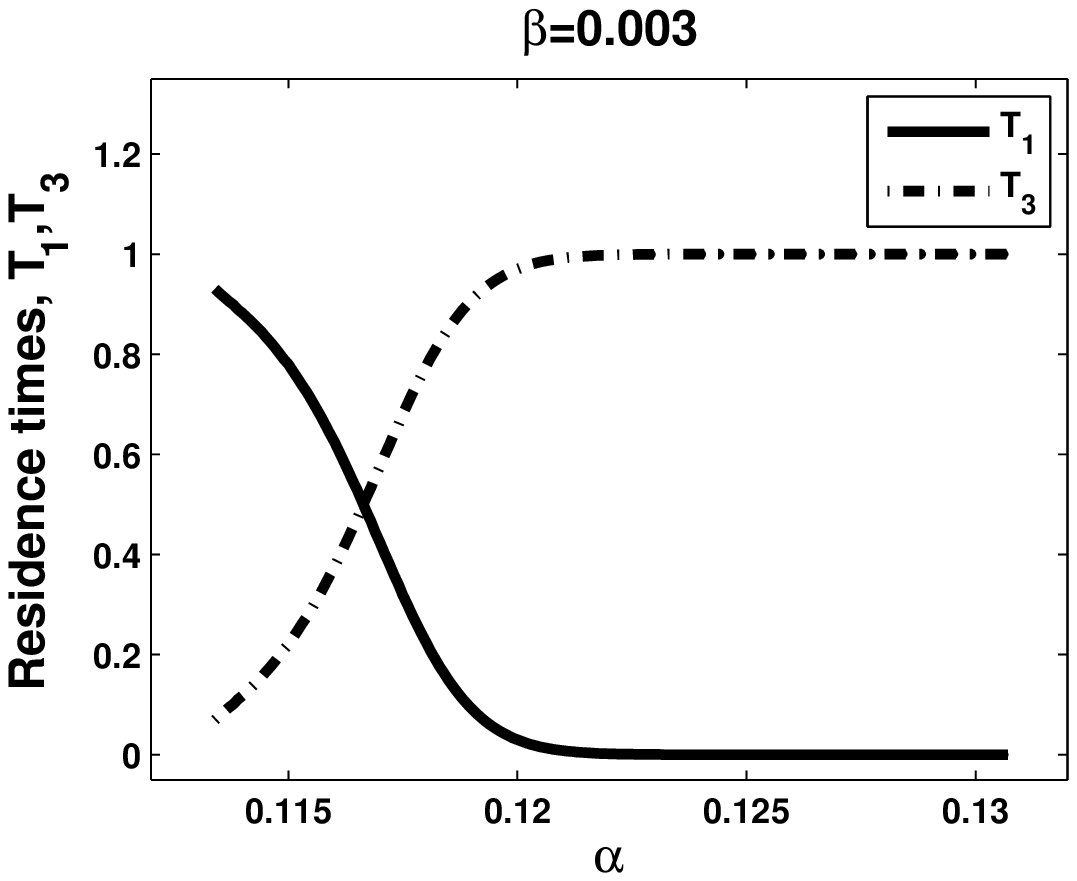,width=10cm,height=5cm,angle=0.0}}
\end{picture}
\caption[] {\footnotesize \it Residence times as a function of the parameter
$\alpha$ for different values of the parameter $\beta$. The noise level is
$D=0.1$,, the nonlinearity $\mu=0.1$ } \label{fig3a}
\end{center}
\end{minipage}
\end{figure}

\begin{figure}[htb]
\centering
\begin{minipage}[10,10]{15cm}
\begin{center}
\begin{picture}(200,150)
\put(20,110){\psfig{file=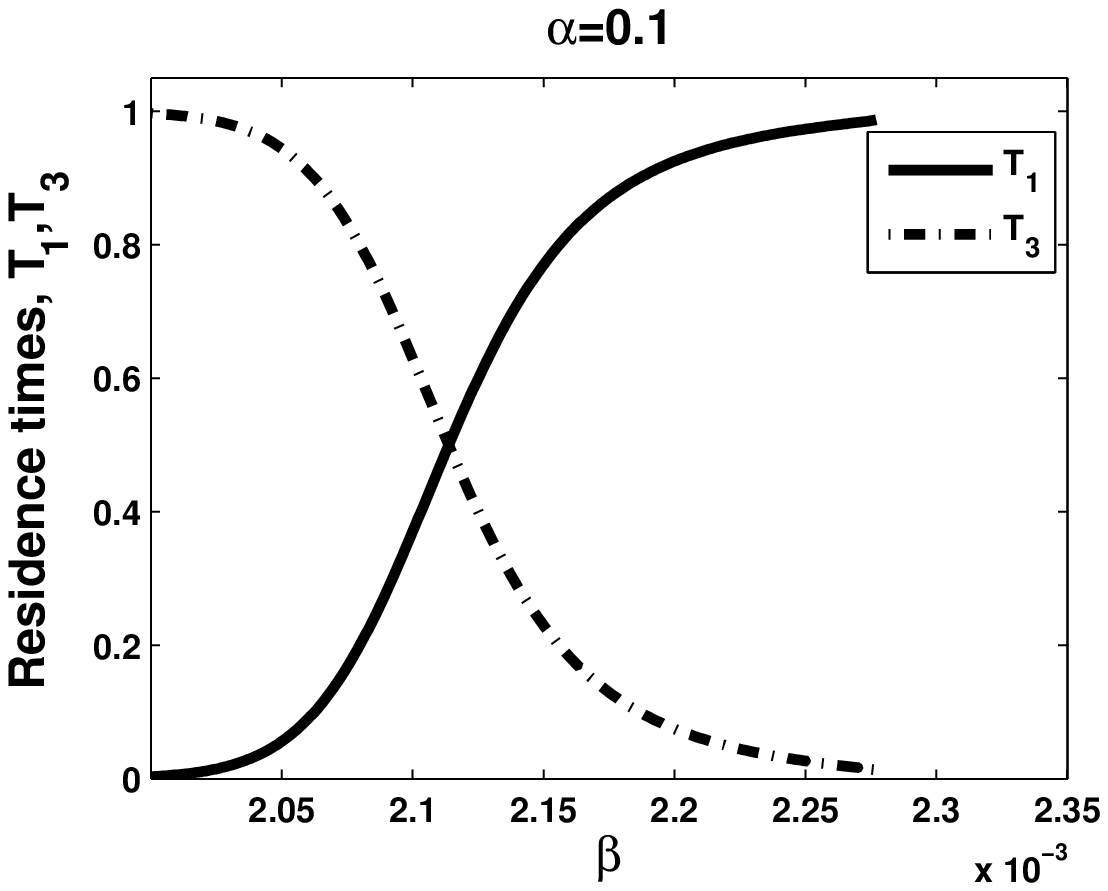,width=10cm,height=5cm,angle=0.0}}
\put(20,55){\psfig{file=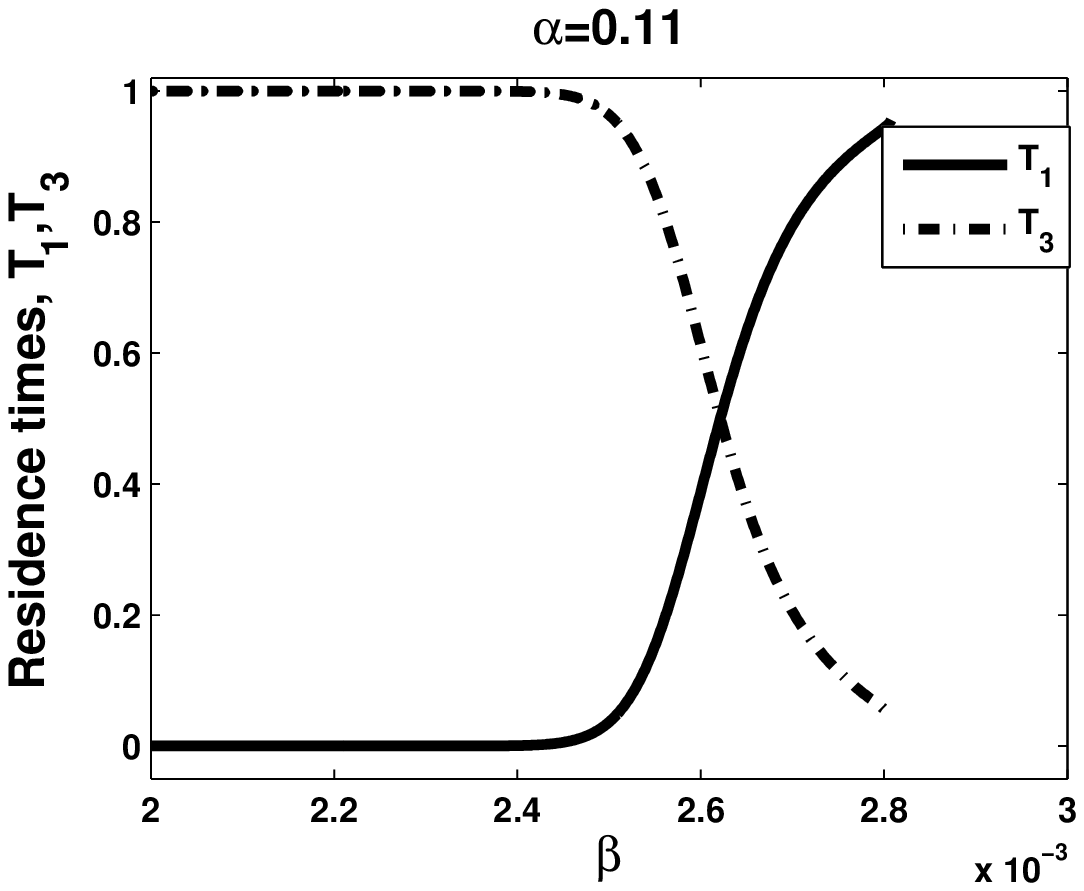,width=10cm,height=5cm,angle=0.0}}
\put(20,0){\psfig{file=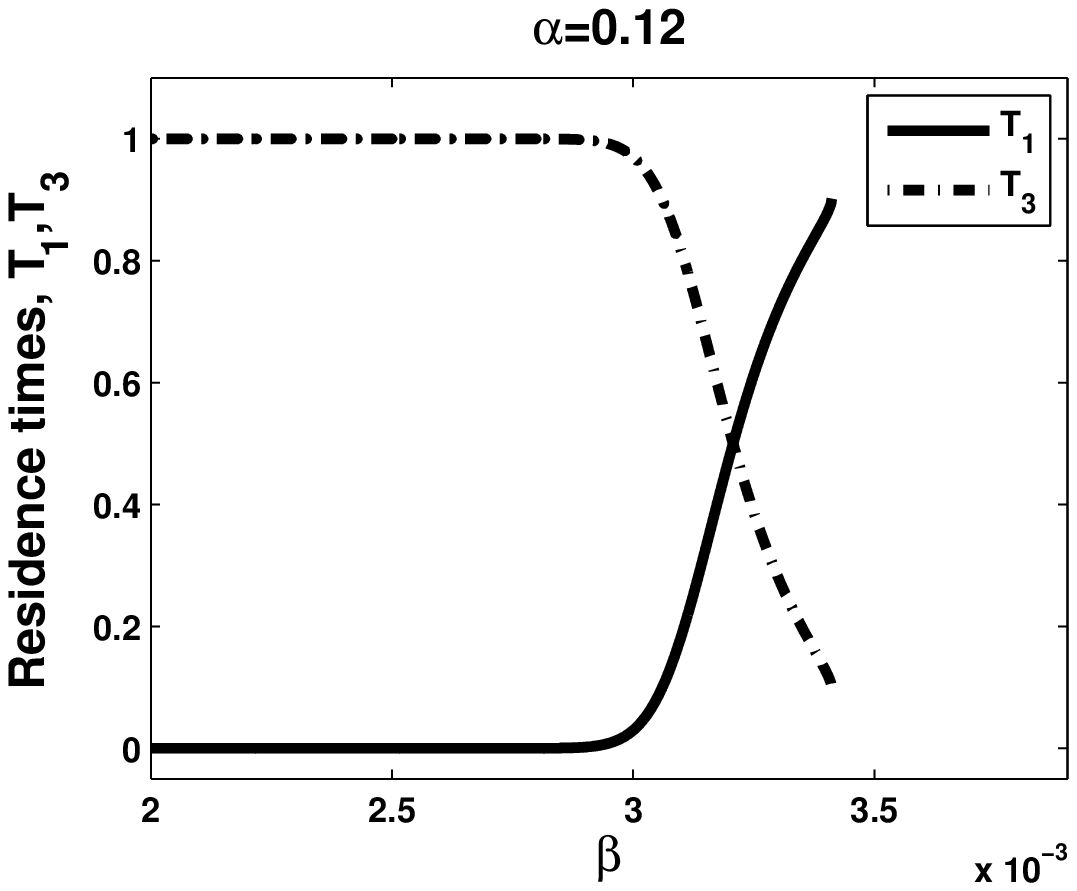,width=10cm,height=5cm,angle=0.0}}
\end{picture}
\caption[] {\footnotesize \it Residence times as a function of the parameter
$\beta$ for different values of the parameter $\alpha$. The noise level is
$D=0.1$, the nonlinearity $\mu=0.1$} \label{fig4a}
\end{center}
\end{minipage}
\end{figure}

\begin{figure}[htb]
\centering
\begin{minipage}[10,10]{15cm}
\begin{center}
\begin{picture}(200,150)
\put(20,110){\psfig{file=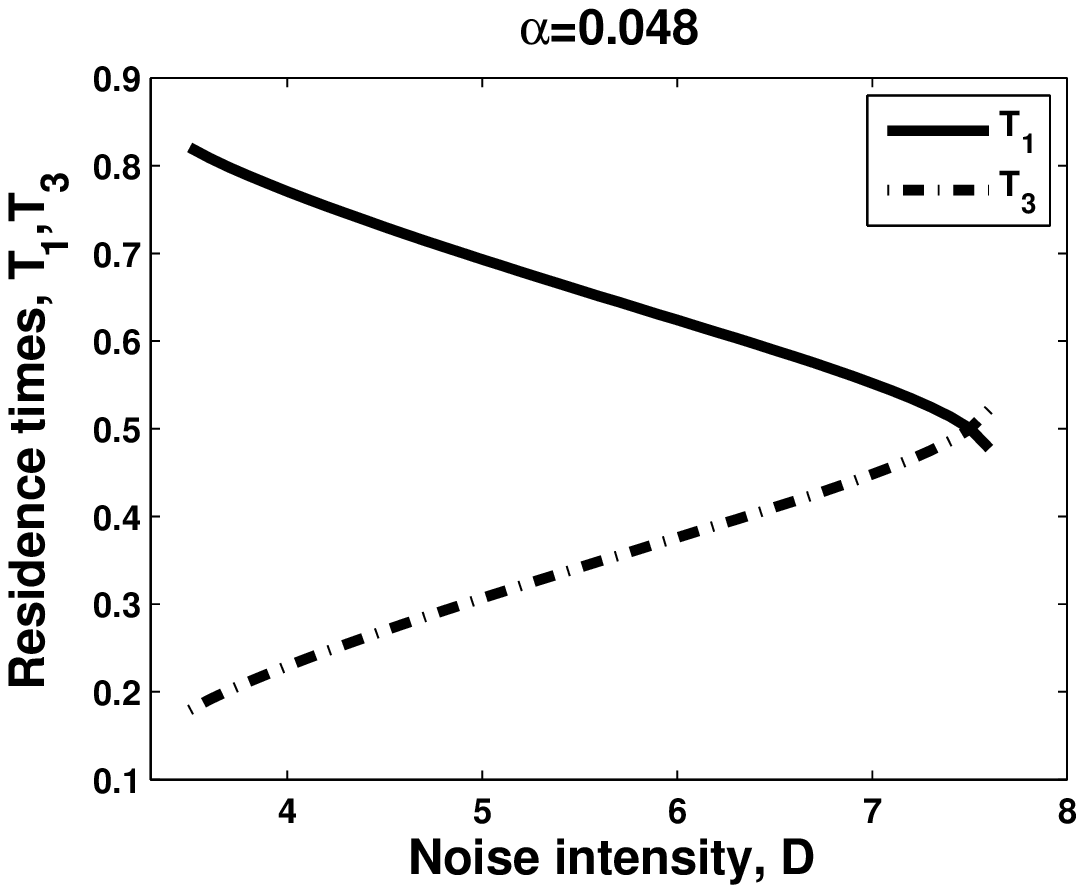,width=10cm,height=5cm,angle=0.0}}
\put(20,55){\psfig{file=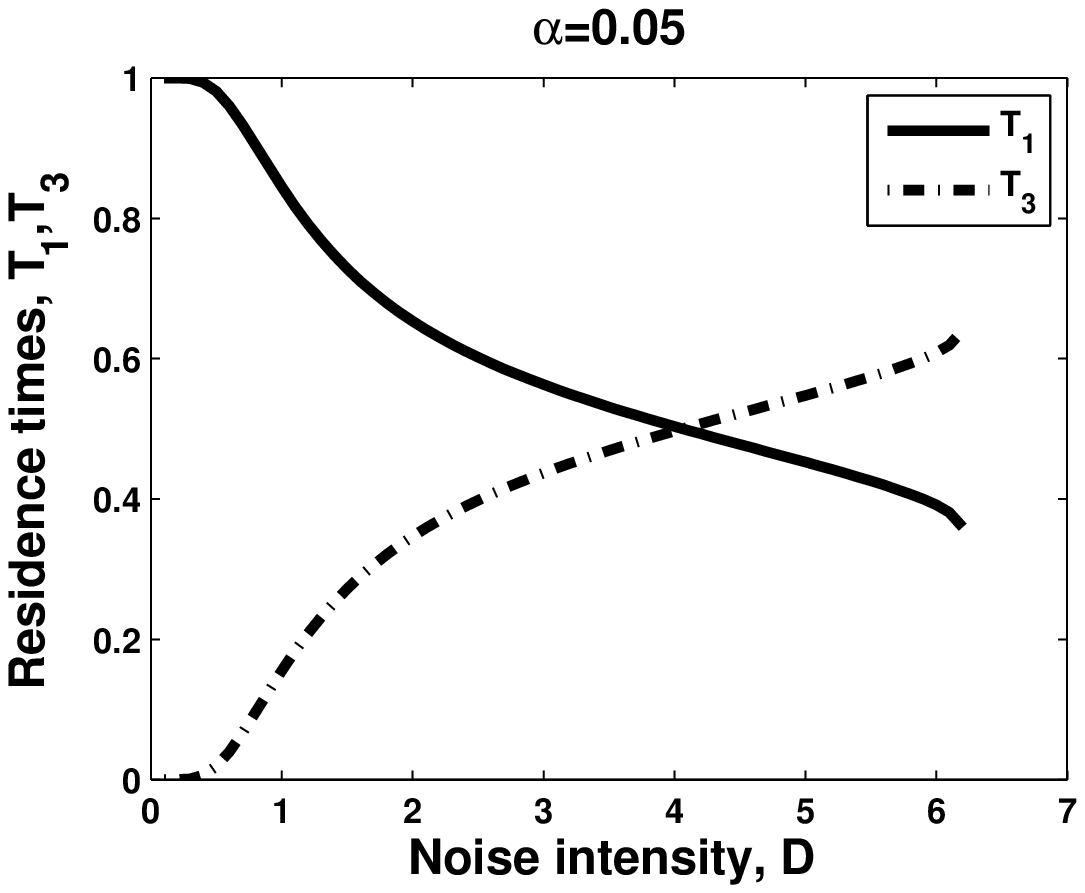,width=10cm,height=5cm,angle=0.0}}
\put(20,0){\psfig{file=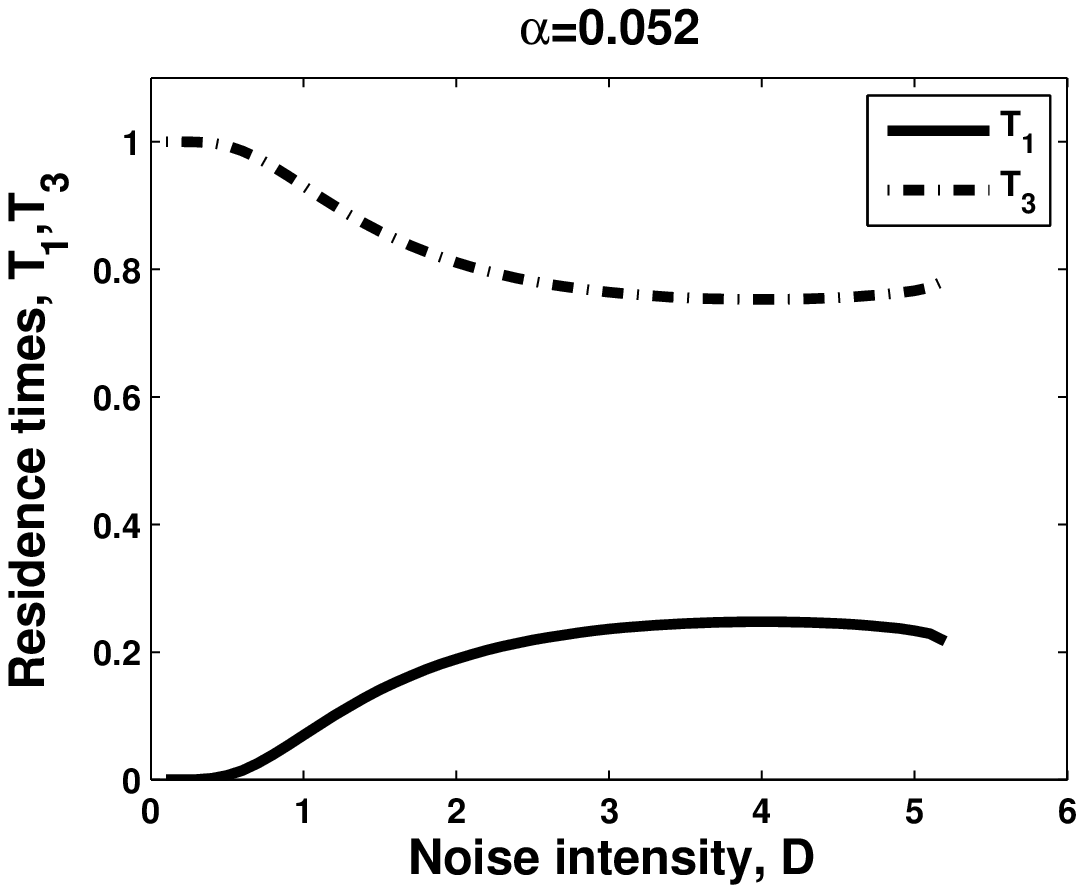,width=10cm,height=5cm,angle=0.0}}
\end{picture}
\caption[] {\footnotesize \it Residence times as a function of the noise
intensity $D$ for different values of the parameter $\alpha$. The second
dissipation parameter reads $\beta=0.0005$, the nonlinearity $\mu=0.1$.}
\label{fig5a}
\end{center}
\end{minipage}
\end{figure}

\begin{figure}[htb]
\centering
\begin{minipage}[10,10]{15cm}
\begin{center}
\begin{picture}(200,100)
\put(0,50){\psfig{file=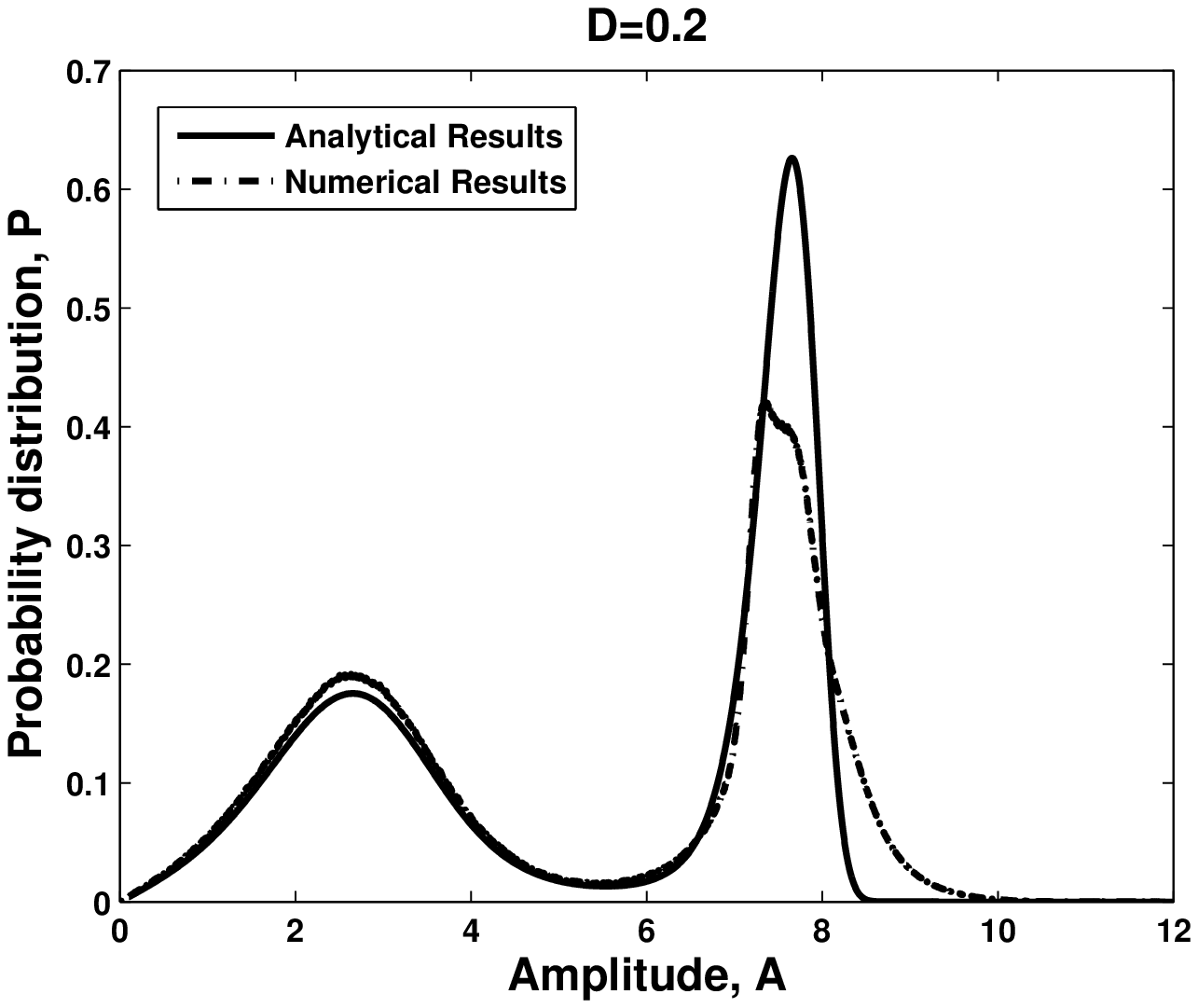,width=8cm,height=5cm,angle=0.0}}
\put(80,50){\psfig{file=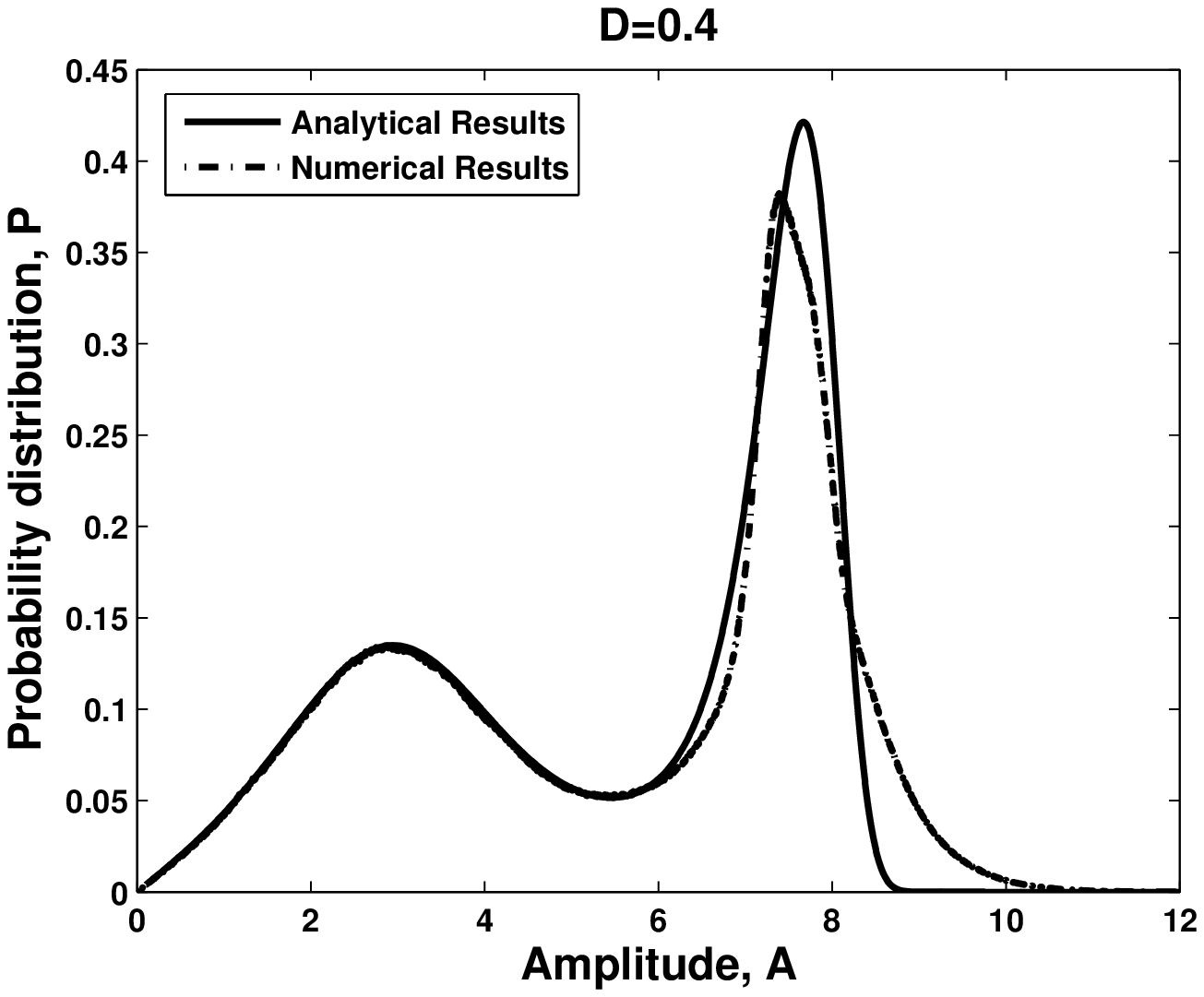,width=8cm,height=5cm,angle=0.0}}
\put(0,0){\psfig{file=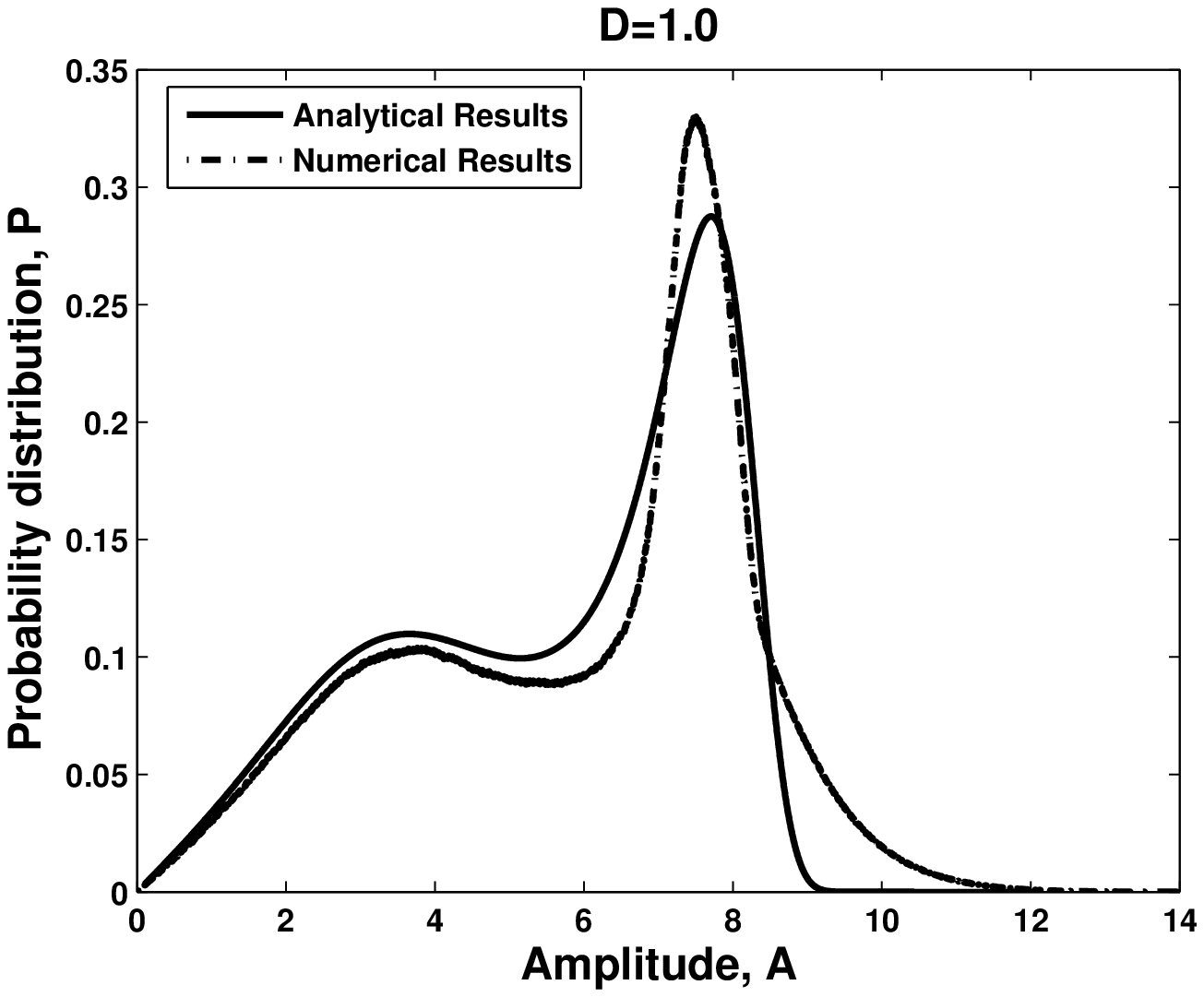,width=8cm,height=5cm,angle=0.0}}
\put(80,0){\psfig{file=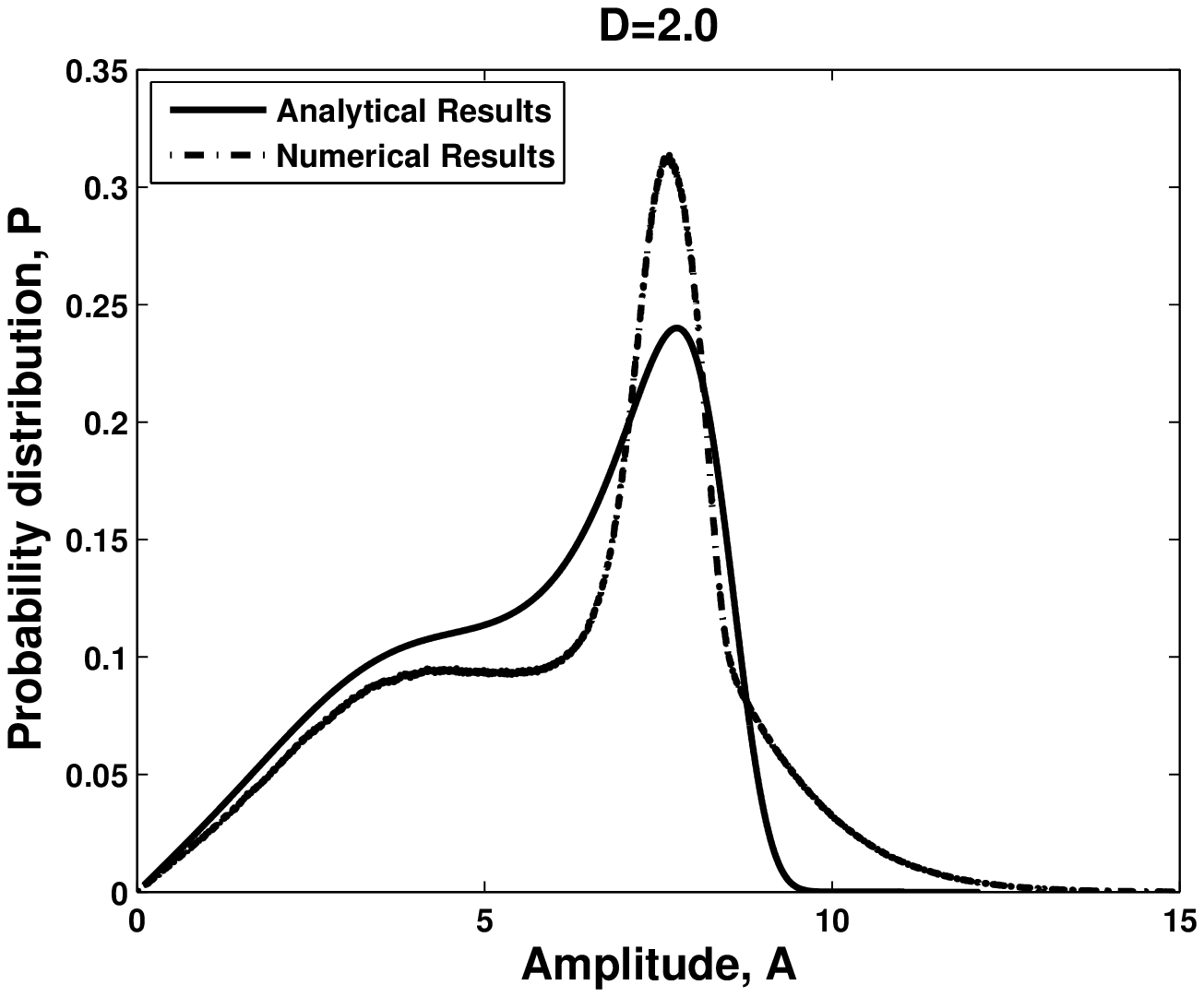,width=8cm,height=5cm,angle=0.0}}
\end{picture}
\caption[] {\footnotesize \it Asymmetric probability distributions for different
values of the noise intensity $D$
versus the amplitude $A$ when the frequencies of both attractors are identical
\emph{i.e} $\Omega_1\simeq
\Omega_3\simeq 1$. Parameters of the system are $\mu=0.1$ and $\alpha=0.083$,
$\beta=0.0014$.} \label{fig4}
\end{center}
\end{minipage}
\end{figure}


\begin{figure}[htb]
\centering
\begin{minipage}[10,10]{15cm}
\begin{center}
\begin{picture}(200,100)
\put(20,50){\psfig{file=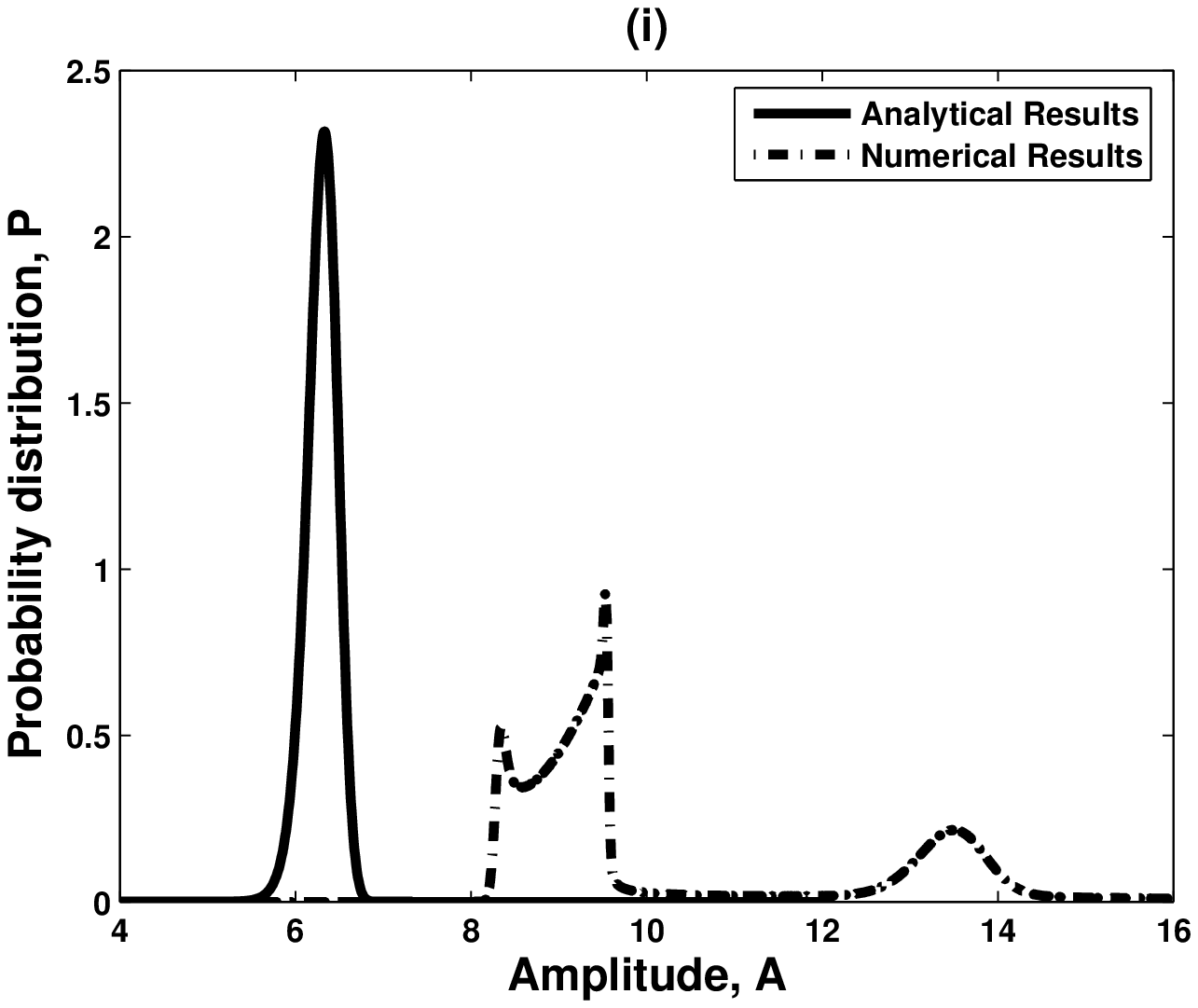,width=10cm,height=5cm,angle=0.0}}
\put(20,0){\psfig{file=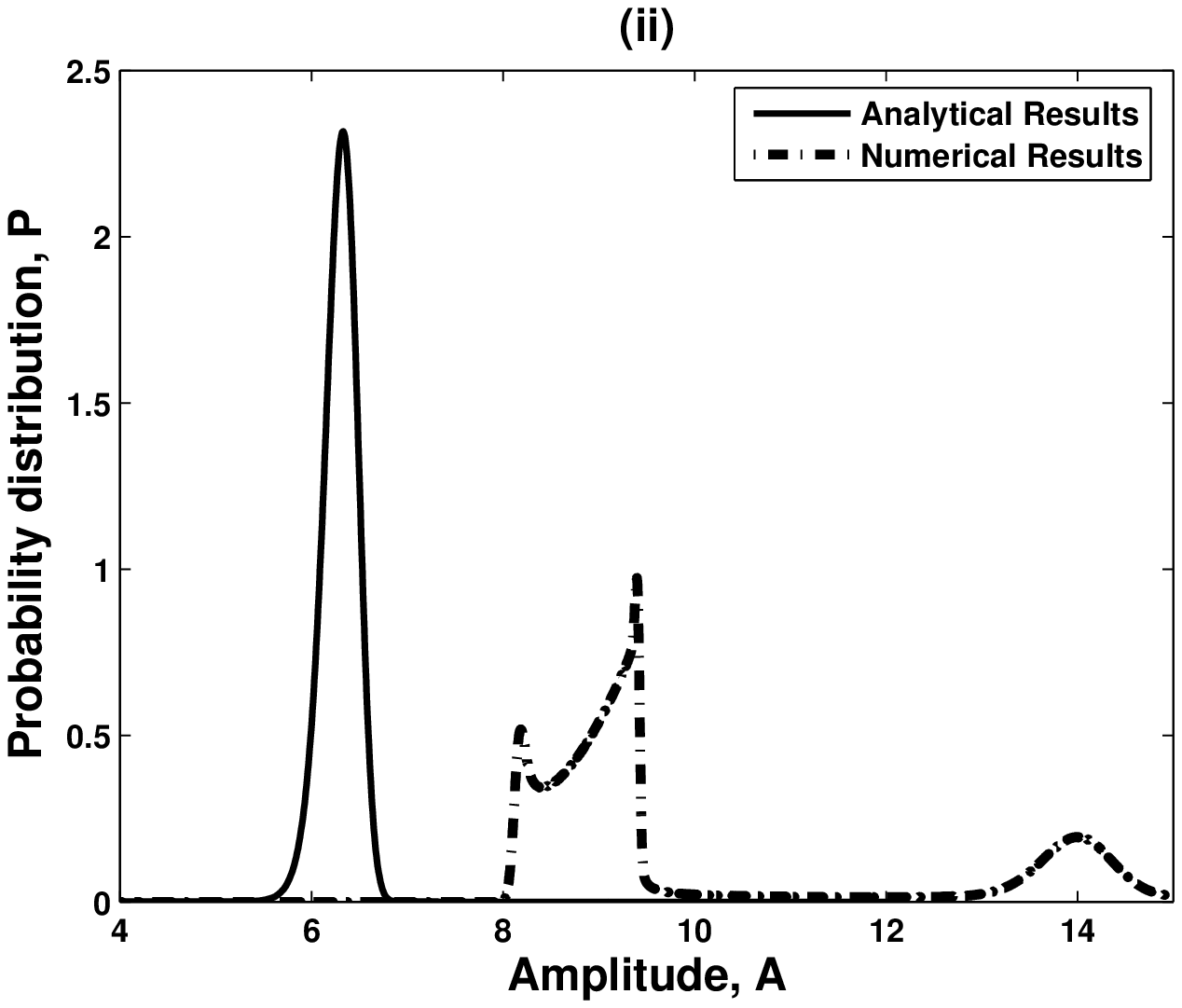,width=10cm,height=5cm,angle=0.0}}
\end{picture}
\caption[] {\footnotesize \it Probability distribution versus
the amplitude $A$ when the frequencies of the attractors are not identical
\emph{i.e} $\Omega_1\neq \Omega_3$.
Parameters of the system are $D=0.1$, $\mu=0.1$, (i): $\alpha=0.09$,
$\beta=0.0012, \Omega_1\simeq 1, \Omega_3\simeq 0.85$ and (ii):
$\alpha=0.1,\beta=0.014,\Omega_1\simeq 1, \Omega_3\simeq 0.8$.} \label{fig5}
\end{center}
\end{minipage}
\end{figure}

\begin{figure}[htb]
\centering
\begin{minipage}[10,10]{15cm}
\begin{center}
\begin{picture}(200,100)
\put(20,50){\psfig{file=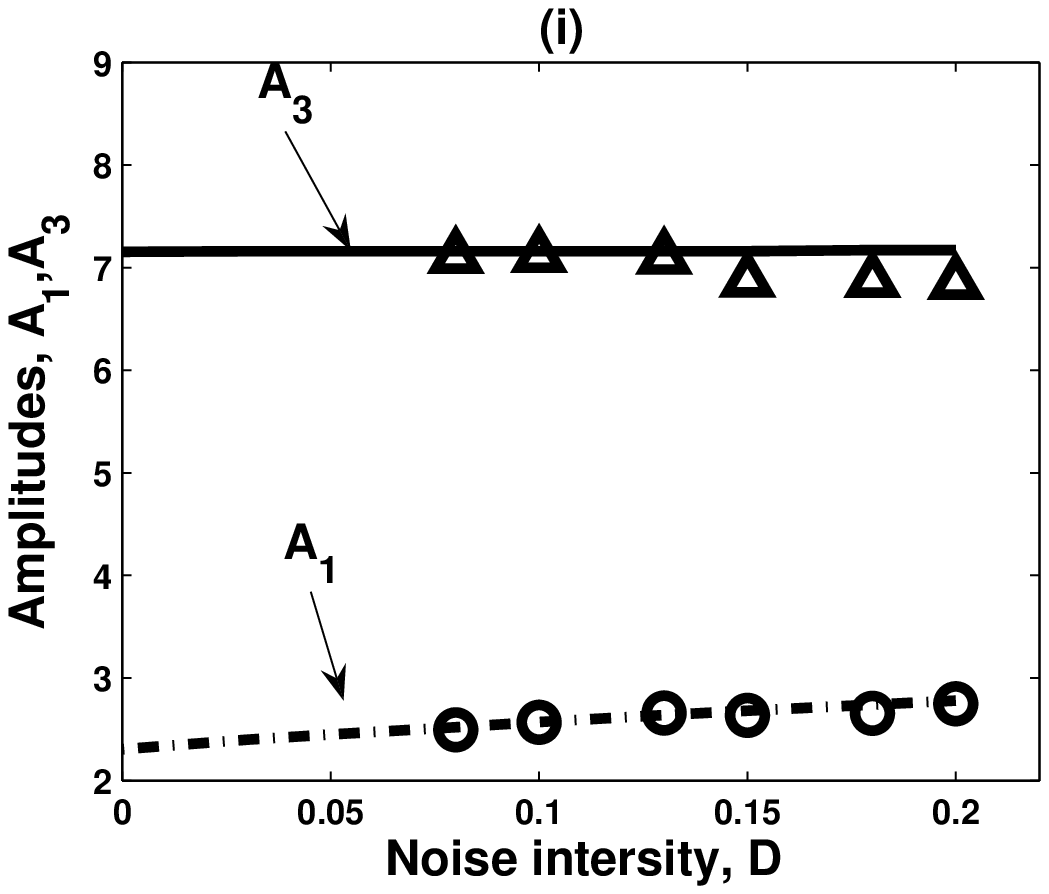,width=10cm,height=5cm,angle=0.0}}
\put(20,0){\psfig{file=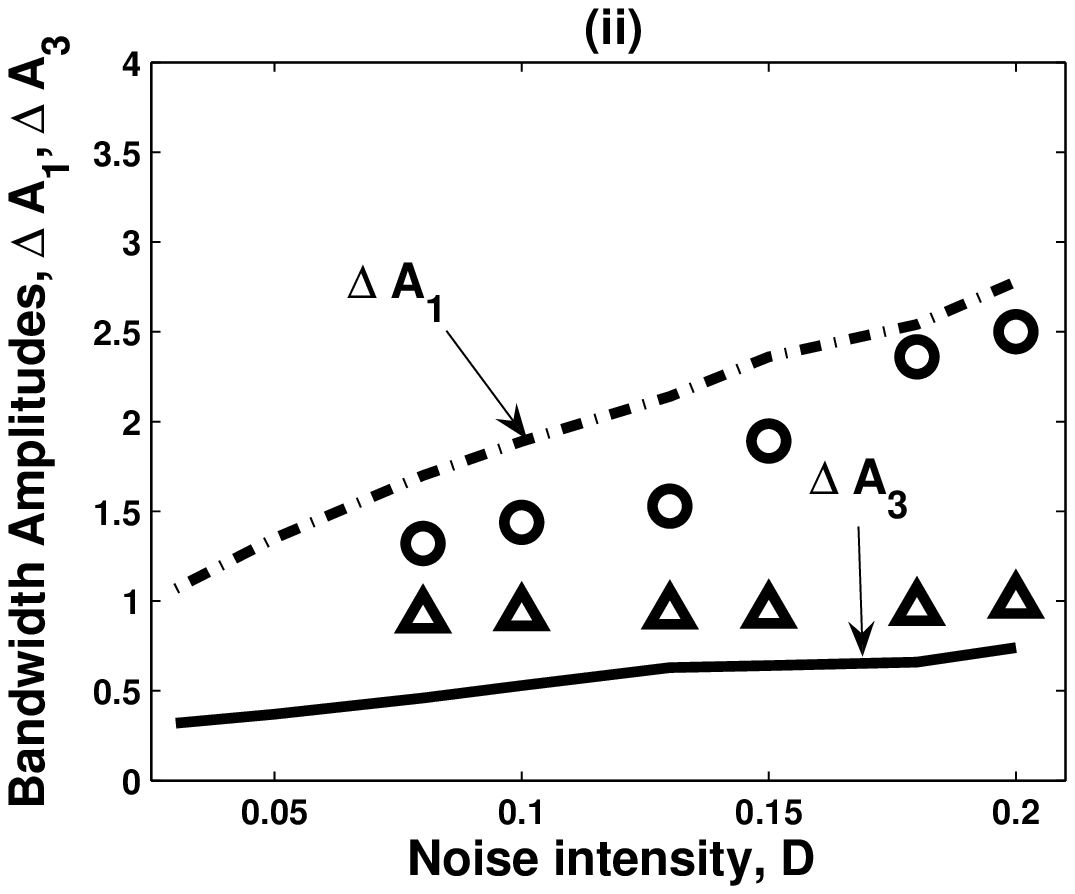,width=10cm,height=5cm,angle=0.0}}
\end{picture}
\caption[] {\footnotesize \it Variation of the amplitudes $A_i$ and the
bandwidths $\Delta A_i$ versus the noise intensity $D$. Lines and symbols denote
analytical and numerical results, respectively. The circles and dot-dashed lines
refer to the inner attractor $A_1$, solid lines and triangles to the outer
attractor $A_3$.  The parameters used are $\mu=0.1$, $\alpha=0.1$,
$\beta=0.002$.} \label{fig6}
\end{center}
\end{minipage}
\end{figure}

\begin{figure}[htb]
\centering
\begin{minipage}[10,10]{15cm}
\begin{center}
\begin{picture}(200,100)
\put(20,50){\psfig{file=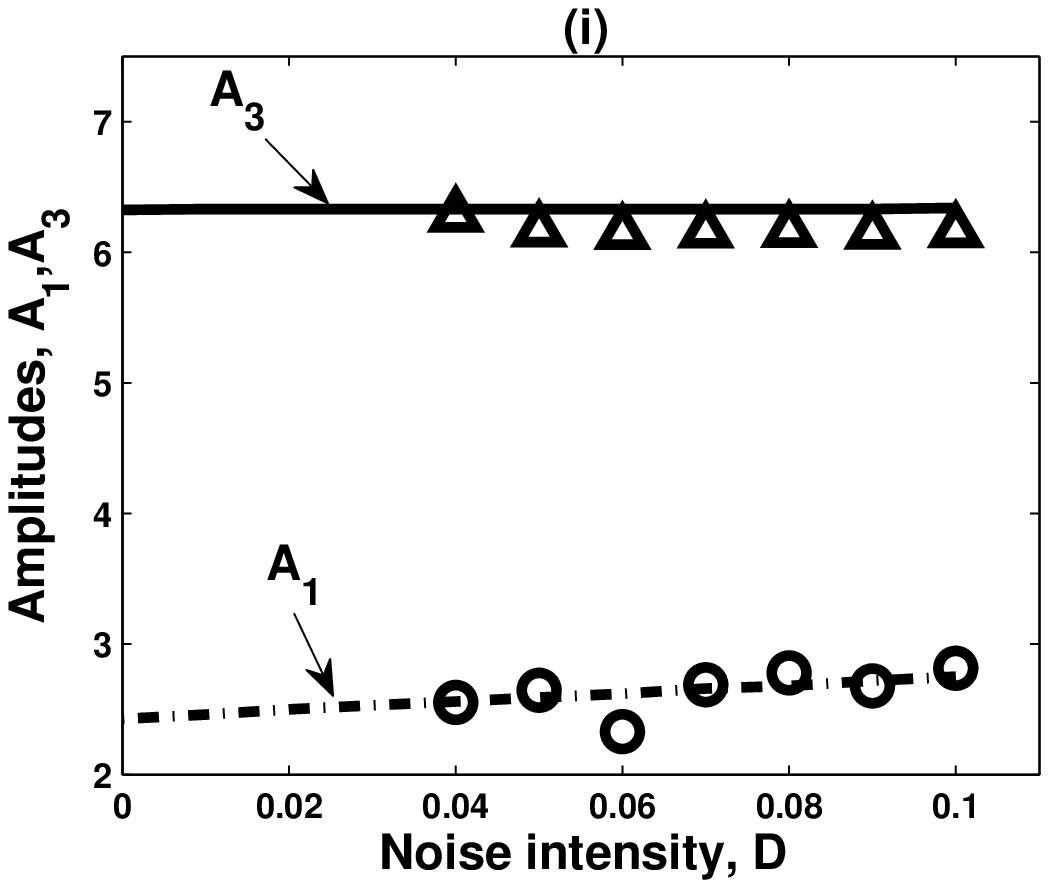,width=10cm,height=5cm,angle=0.0}}
\put(20,0){\psfig{file=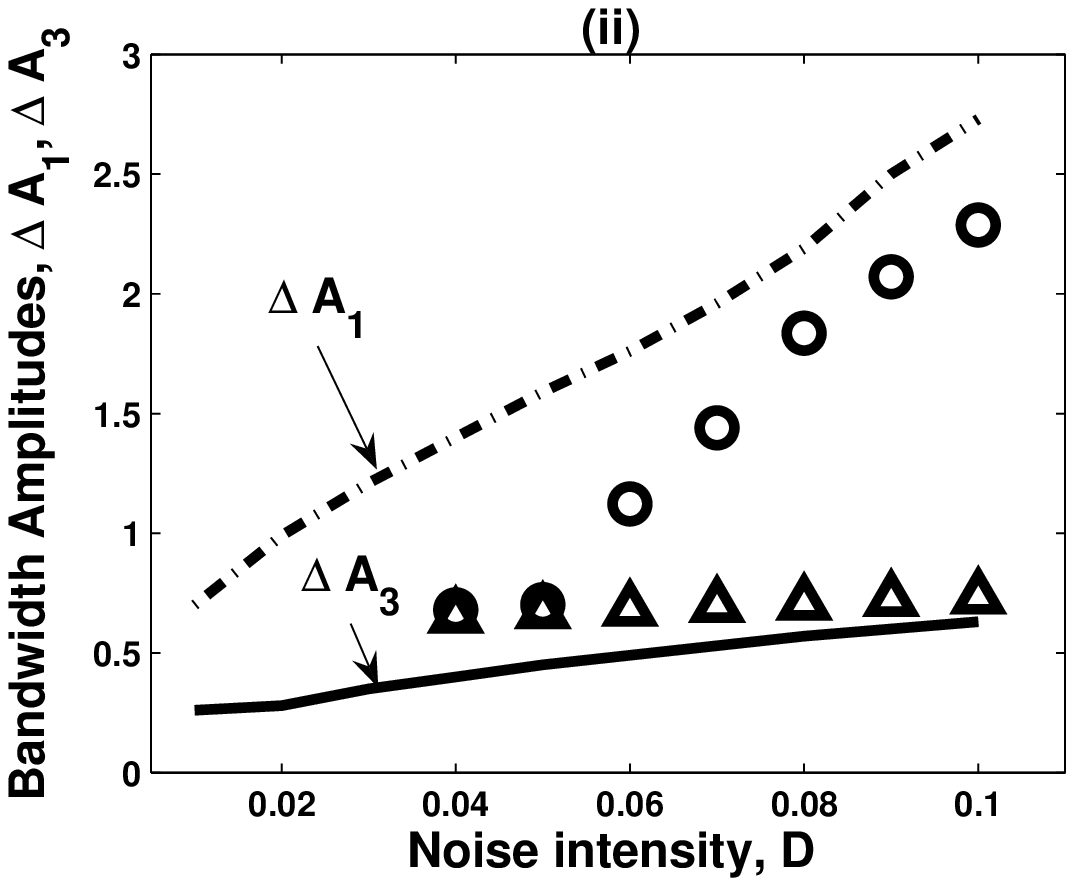,width=10cm,height=5cm,angle=0.0}}
\end{picture}
\caption[] {\footnotesize \it Variation of the amplitudes $A_i$ and the
bandwidths $\Delta A_i$ versus the noise intensity $D$. Lines and symbols denote
analytical and numerical results, respectively. The circles and dot-dashed lines
refer to the inner attractor $A_1$, solid lines and triangles to the outer
attractor $A_3$.  The parameters used are $\mu=0.1$,
 $\alpha=0.12$, $\beta=0.003$.} \label{fig7}
\end{center}
\end{minipage}
\end{figure}

\begin{figure}[htb]
\centering
\begin{minipage}[10,10]{15cm}
\begin{center}
\begin{picture}(200,100)
\put(20,50){\psfig{file=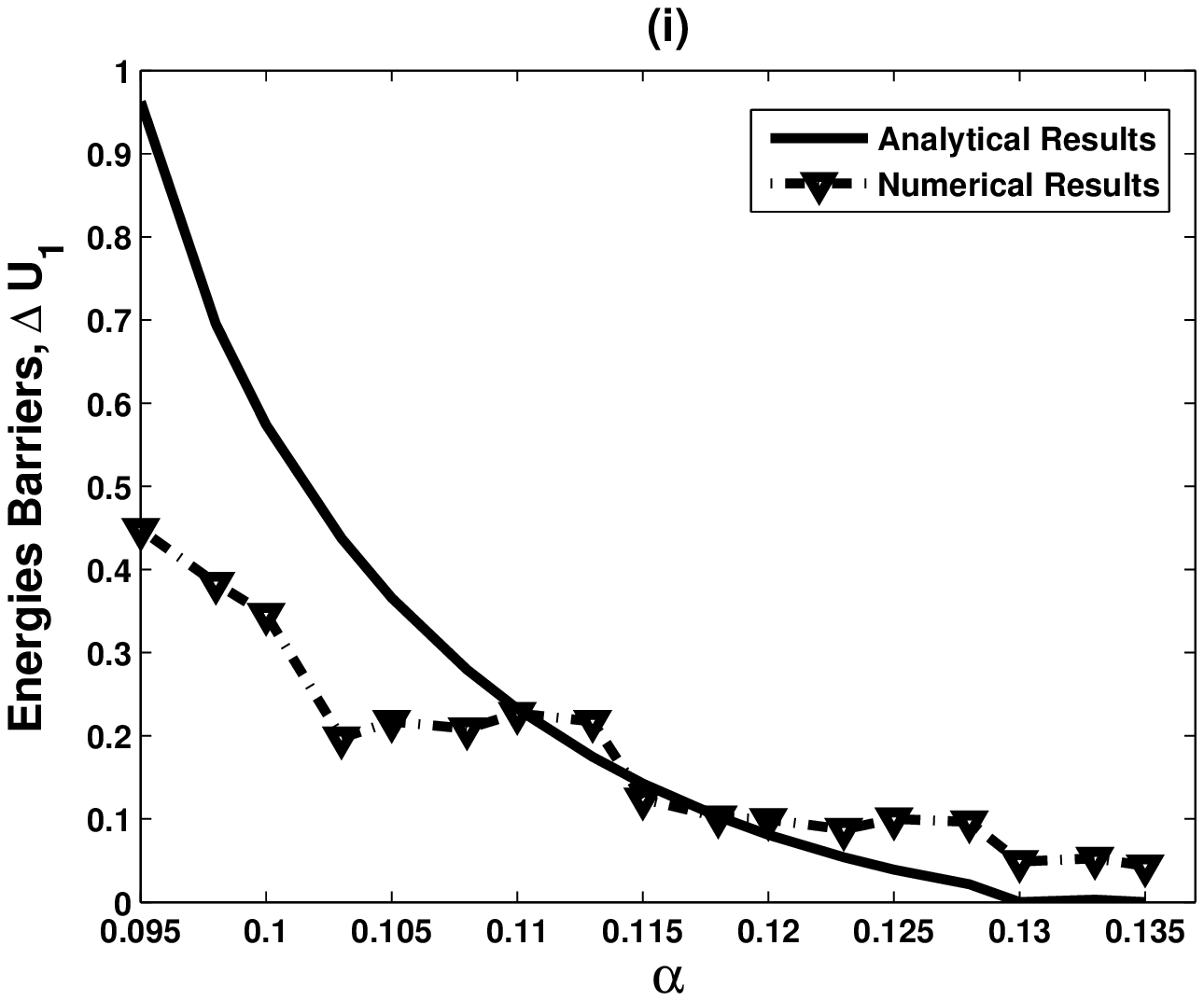,width=10cm,height=5cm,angle=0.0}}
\put(20,0){\psfig{file=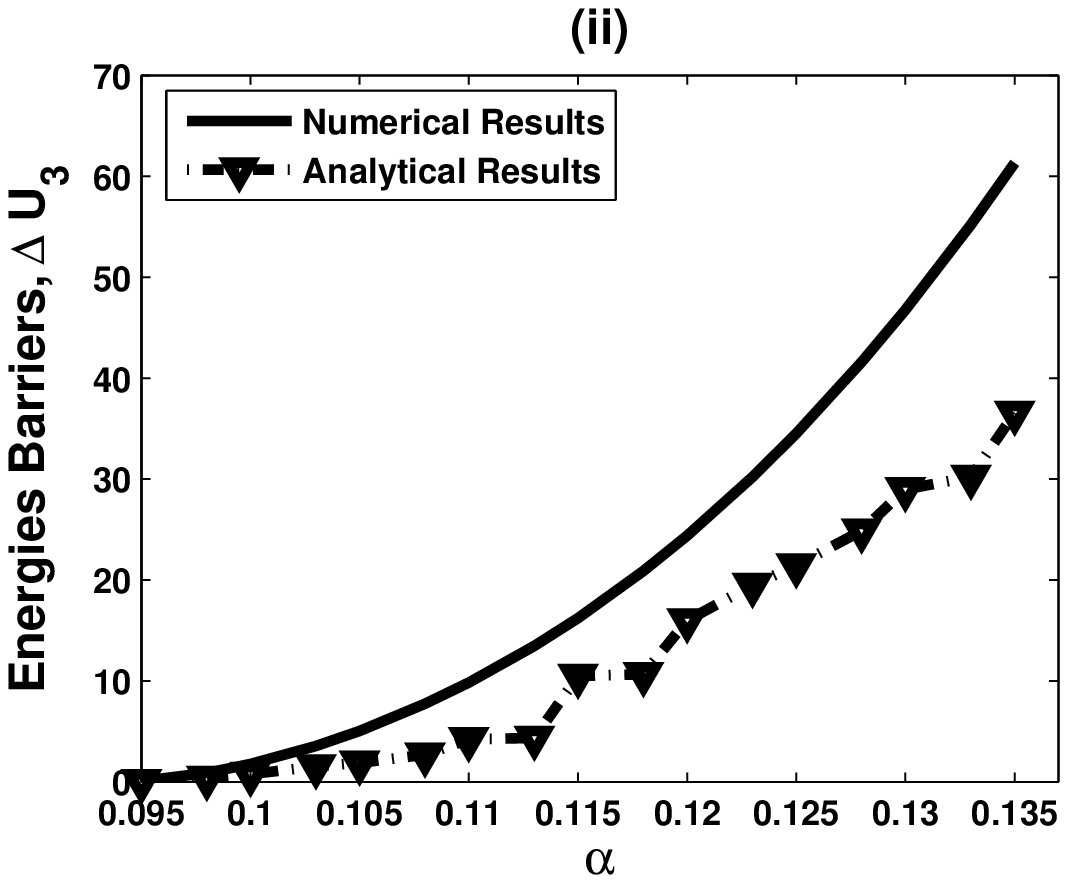,width=10cm,height=5cm,angle=0.0}}
\end{picture}
\caption[] {\footnotesize \it Behavior of energy barriers
versus $\alpha$. Solid lines denote the analytical results, while dashed
lines with triangles denote numerical simulations. Parameters of the system are
$\mu=0.1$ and
$\beta=0.002$. } \label{fig8}
\end{center}
\end{minipage}
\end{figure}

\begin{figure}[htb]
\centering
\begin{minipage}[10,10]{15cm}
\begin{center}
\begin{picture}(200,100)
\put(20,50){\psfig{file=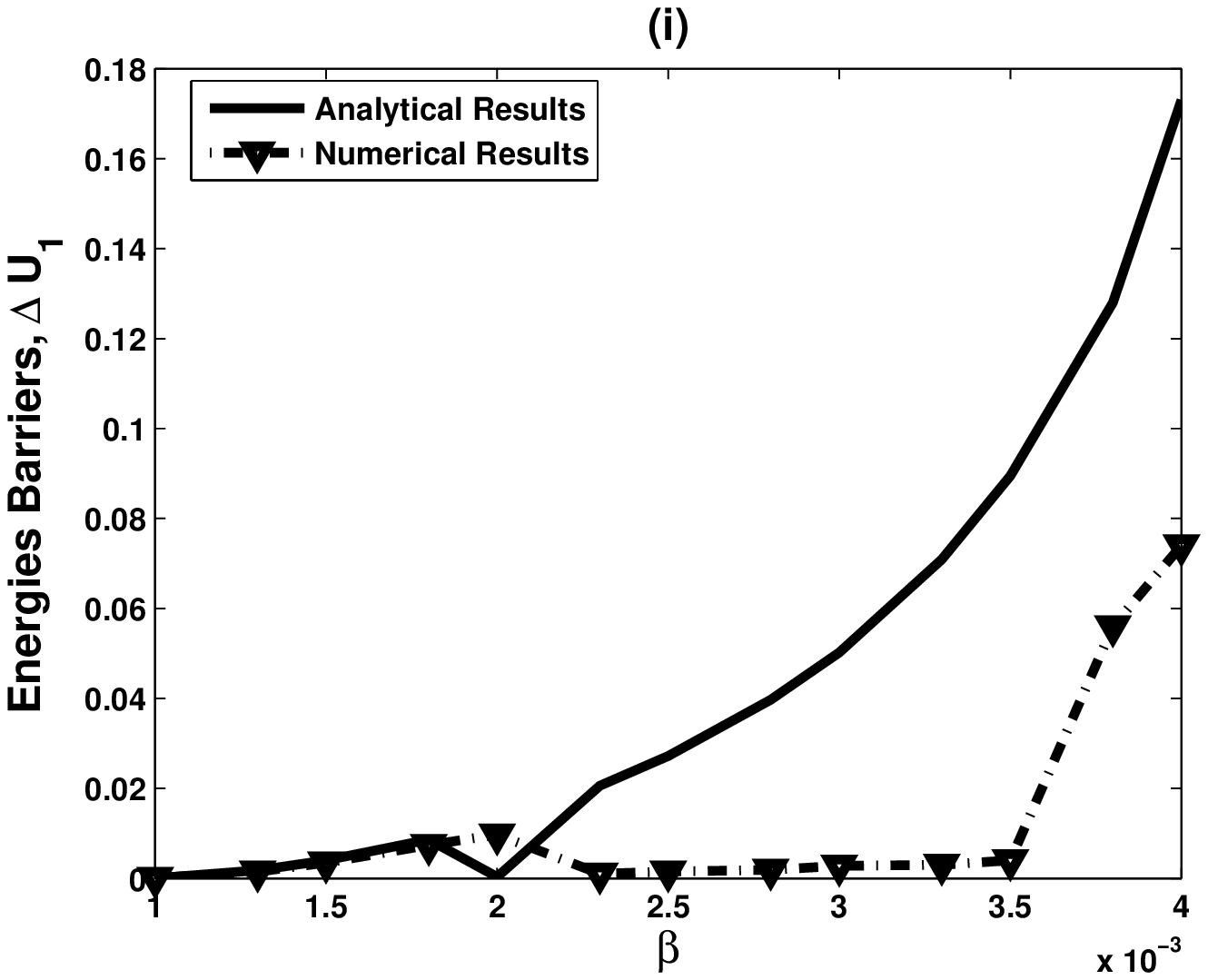,width=10cm,height=5cm,angle=0.0}}
\put(20,0){\psfig{file=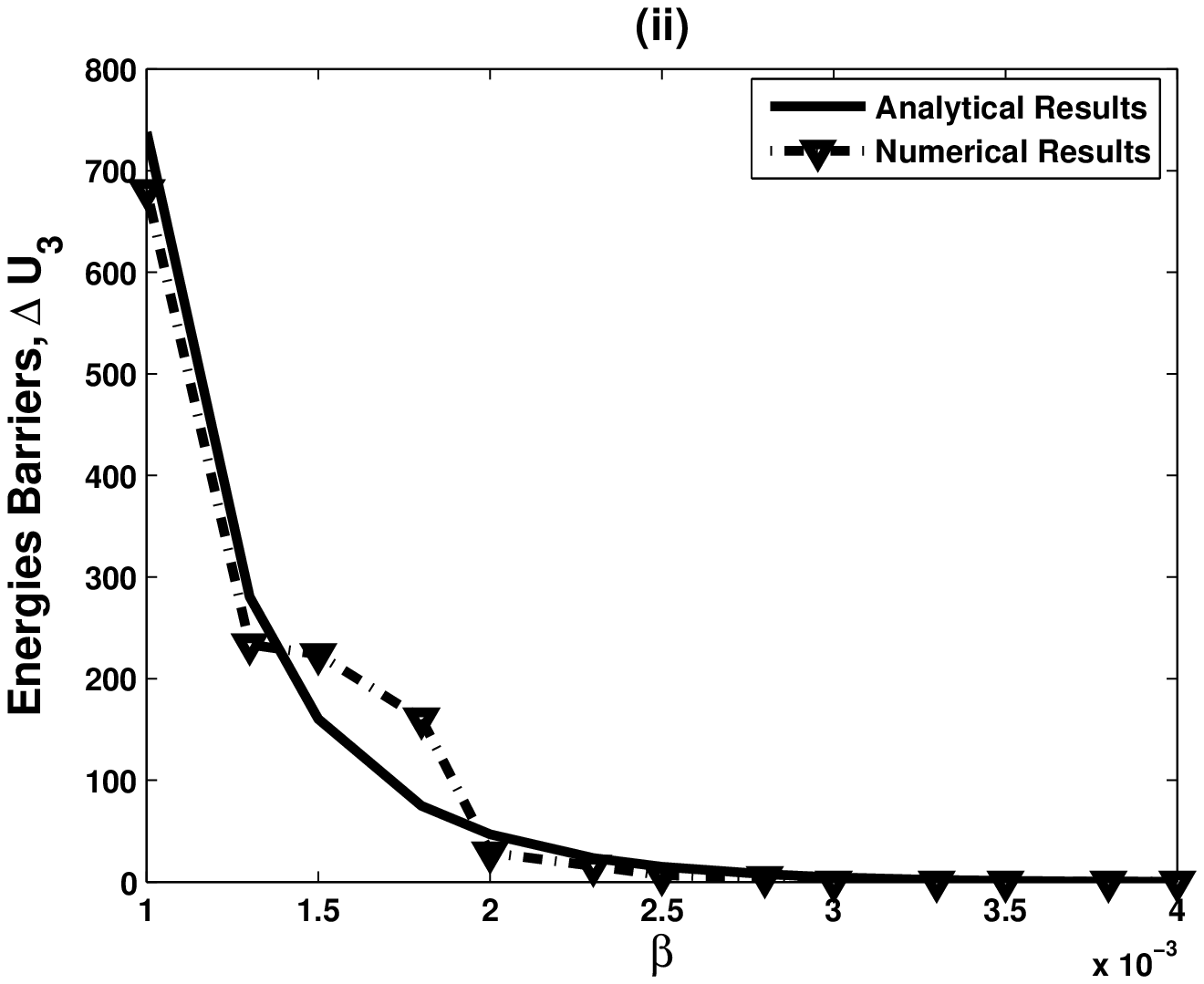,width=10cm,height=5cm,angle=0.0}}
\end{picture}
\caption[] {\footnotesize \it Behavior of energy barriers
versus $\beta$.
Solid lines denote the analytical results, while dashed
lines with triangles denote numerical simulations. Parameters of the system are
$\mu=0.1$ and
$\alpha=0.13$. } \label{fig9}
\end{center}
\end{minipage}
\end{figure}

\end{document}